%% file: arx.tex
\documentclass[11pt,reqno]{amsart}      

\usepackage{amssymb,amsthm}             

\usepackage{eucal}                      
\newcommand{\EUcal}{\mathcal}		

\usepackage{calc}			
\usepackage{graphicx}
\usepackage{epic,eepic}			

\numberwithin{equation}{section}        
\newcommand{\mnote}[1]{} 

\renewcommand{\Re}{{\mathbb R}}         
\newcommand{\la}{\langle}               
\newcommand{\ra}{\rangle}               
\newcommand{\half}{\frac{1}{2}}         
\newcommand{\third}{\frac{1}{3}}        
\newcommand{\quart}{\frac{1}{4}}        
\newcommand{\agamma}{{\bar \gamma}}     
\newcommand{\tens}{\otimes}		
\newcommand{\Teich}{\CMcal T}		
\newcommand{\Energy}{\mathcal E}        

\newcommand{\U}{\text{\rm U}}           
\newcommand{\SO}{\text{\rm SO}}         

\newcommand{\EBR}{{\CMcal Q}}         

\newcommand{\DD}{\CMcal D}            

\newcommand{\aM}{{\bar M}}              
\newcommand{\anabla}{{\bar \nabla}}     
\newcommand{\ame}{{\bar g}}             
\newcommand{\aGamma}{{\bar \Gamma}}      
\newcommand{\aR}{{\bar R}}              
\newcommand{\tM}{\tilde M}
\newcommand{\tme}{\tilde g}		
\newcommand{\Scri}{\EUcal I}	

\newcommand{\TT}{\text{\sc TT}}         
\newcommand{\Vol}{\text{\rm Vol}}       
\newcommand{\bfi}{\text{\bf i}}         
\newcommand{\Eps}{\epsilon}             
\newcommand{\eps}{\epsilon}             
\newcommand{\Lie}{{\CMcal L}}         
\newcommand{\tr}{{\text{\rm tr}}}       
\newcommand{\diag}{{\text{\rm diag}}}   
\renewcommand{\div}{\text{\rm div}}     
\newcommand{\Lapse}{N}                  
\newcommand{\hme}{\hat g}		
\newcommand{\hGamma}{\hat \Gamma}	
\newcommand{\Shift}{X}                  
\newcommand{\Id}{\text{Id}}		

\newcommand{\bfH}{\mathbf H}
\newcommand{\bfE}{\mathbf E}

\newcommand{\tSL}{\widetilde{\text{SL}}}
\newcommand{\frakg}{\mathfrak g}

\newcommand{\Sigp}{\Sigma_{+}}
\newcommand{\Sigm}{\Sigma_{-}}
\newcommand{\Sigc}{\Sigma_{\times}}

\def\Nm{N_{-}}
\def\Nc{N_{\times}}
\newcommand{\vhat}{{\hat v}}
\newcommand{\vhyp}{v_{\bfH^2}}
\newcommand{\vhathyp}{\vhat_{\bfH^2}}
\newcommand{\vsign}{v}

\theoremstyle{plain}
\newtheorem{thm}{Theorem}[section]

\newtheorem{conj}{Conjecture}

\newtheorem{example}[thm]{Example}
\newtheorem{definition}[thm]{Definition}
\newtheorem{prop}[thm]{Proposition}
\newtheorem{remark}{Remark}[section]

\title{The global existence problem in general relativity}
\author[L. Andersson]{Lars Andersson$^1$}
\thanks{$^1$Supported in part by the Swedish Natural
Sciences Research Council,  contract no.  F-FU 4873-307, and
the NSF under
contract no. DMS 0104402.
}
\address{Department of Mathematics\\
University of Miami\\
Coral Gables, FL 33124\\
USA}
\email{larsa\char'100math.miami.edu}
\date{June 27, 2004}

\begin{document}
\begin{abstract}
We survey some known facts and open questions concerning the global properties 
of 3+1 dimensional spacetimes containing a compact Cauchy surface. 
We consider 
spacetimes with an $\ell$--dimensional Lie algebra of space--like Killing
fields. 
For each $\ell \leq 3$, we give some basic results and 
conjectures on global existence and
cosmic censorship. 
\end{abstract}
\maketitle

\tableofcontents
\section{Introduction}
In this review, we will describe some results and conjectures about the global
structure of solutions to the Einstein equations in 3+1 dimensions. 
We consider spacetimes $(\aM, \ame)$ with an $\ell$--dimensional 
Lie algebra of space--like Killing fields. We may say that 
such spacetimes have
a (local) isometry group $G$ of 
dimension $\ell$ with the action
of $G$ generated by space--like Killing fields.

For each value $\ell \leq 3$ of the dimension of the isometry group, 
we state the reduced field equations as well as attempt to give an overview 
of the most important results and conjectures. We will concentrate on the
vacuum case. 

In section \ref{sec:einst}, we present the Einstein equations and give the
3+1 decomposition into constraint and evolution equations, cf. 
subsection \ref{sec:evoleqs}. Due to the gauge
freedom in the Einstein equations, questions on the global properties of
solutions to the Einstein equations must be posed carefully. We introduce
the notions of vacuum extension and maximal Cauchy development and state the
uniqueness theorem of Choquet--Bruhat and Geroch, for maximal vacuum Cauchy
developments. We also collect here a few basic facts about Killing fields on
globally hyperbolic spacetimes. In subsection \ref{sec:SCC} a 
version of the cosmic censorship conjecture appropriate
for vacuum spacetimes, with compact Cauchy surface, is stated and in 
subsection \ref{sec:CMC} we
discuss a few gauge conditions which may be of use for the global existence
problem for the Einstein equations. 
Section \ref{sec:einst} is ended by a discussion of  a few PDE aspects of
the Einstein equations which are relevant for the topic of this paper, cf.
subsection \ref{sec:PDE}. 

In the cases $\ell = 3$ (Bianchi, cf. section \ref{sec:bianchi}) and  a 
special case of $\ell=2$ 
(polarized Gowdy, cf. section \ref{sec:gowdy}), the global behavior of the
Einstein equations is well understood, both with regard to
global existence of the evolution equations and the cosmic censorship
problem. For the general $\ell=2$ case (local $\U(1)\times \U(1)$ symmetry, 
cf. section \ref{sec:gowdy}), there are only partial results on the global
existence problem and the cosmic censorship problem remains open,
although conjectures supported by numerical evidence give a 
good indication of what the correct 
picture is.
\mnote{check how much of global existence is known for $\U(1)\times
\U(1)$. What about Vinces ``corners''??}

In the cases $\ell = 1$ ($\U(1)$ symmetry, cf. section \ref{sec:U(1)}) 
and $\ell=0$ (no symmetry, i.e. full 3+1 dimensional Einstein equations, cf.
section \ref{sec:3+1}), the large data global existence and cosmic 
censorship problems are open. In the $\U(1)$ case conjectures supported by
numerical evidence give a good idea of the generic behavior, and there is a
small data semi--global existence result 
for the expanding direction due to Choquet--Bruhat and Moncrief
\cite{ChBM:U(1):global,ChB:U(1):global}. 

For 3+1 Einstein gravity without symmetries the only global existence results
known are the theorem on nonlinear stability of Minkowski space of 
Christodoulou and Klainerman, the semi--global existence theorem 
of Friedrich for the hyperboloidal initial value proble and the semi--global
existence theorem for spatially compact spacetimes with Cauchy surface of
hyperbolic type, due to Andersson and Moncrief. These are all
small data results, see section \ref{sec:3+1} for discussion. 

Due to the high degree of complexity of the numerical solution of the
Einstein equations in 3+1 dimensions it is too early to draw any
conclusions relevant to the asymptotic behavior at the singularity for the
full 3+1 dimensional Einstein equations, from the numerical studies 
being performed. However, an attractive scenario is given by the so called
BKL picture, cf. section \ref{sec:concluding} for some remarks and references. 

The Einstein equations are derived from a variational principle, and can be
formulated as a Hamiltonian system (or time--dependent Hamiltonian system,
depending on the gauge), and therefore the Hamiltonian aspect of the
dynamics should not be ignored, see eg. the work by Fischer and Moncrief
on the Hamiltonian reduction of the Einstein equations,
\cite{fischer:moncrief:hamred} and references therein. In fact, 
the Hamiltonian point of view on
the Einstein equations has played a vital role as motivation and guide in
the development of the results discussed here. 
The notion that the Einstein
evolution equations in terms of canonical variables can be viewed as the 
geodesic spray for a metric on the phase space (deWitt metric) modified by a
curvature potential, is natural from the Hamiltonian point of view, 
and this picture has been relevant to the development of 
ideas on asymptotic velocity dominance, see sections \ref{sec:gowdy} and
\ref{sec:U(1)}. 

In this review, however, we will concentrate exclusively on the differential
geometric and analytical point of view. Even with this restriction, many
important topics are left out and we make no claim of complete coverage.
See also \cite{rendall:living2002} and 
\cite{klainerman:nicolo:review} for related surveys.

\noindent{\bf Acknowledgements:} I am 
grateful to Vince Moncrief for 
numerous conversations on the topics covered here 
and for detailed comments on an early version. 
Thanks are due to H{\aa}kan Andr\'easson, Piotr Chru\'sciel, 
Jim Isenberg and Alan Rendall and others
for helpful comments. I am happy to acknowledge the hospitality and support of 
the Institute of
Theoretical Physics, UCSB, 
and Institut des Hautes \'Etudes Scientifiques, Bures sur Yvette, where where part of the writing
was done. 

\section{The Einstein equations}\label{sec:einst}
Let $(\aM, \ame)$ be a smooth 4-dimensional Lorentz 
manifold\footnote{We denote the covariant derivative 
and curvature tensors associated
to $(\aM, \ame)$ by $\anabla, \aR_{abcd}$ etc.
All manifolds are assumed to be Hausdorff, second countable and 
$C^{\infty}$, and all
fields are assumed to be $C^{\infty}$ unless otherwise stated.}
of signature $-$+++. The Lorentzian metric $\ame$ defines a causal structure 
on $\aM$. For the convenience of the reader we give a quick 
review of the basic causality concepts in appendix \ref{sec:causality}, 
see \cite[Chapter
8]{wald:text},\cite{beem:ehrlich:global,hawking:ellis,penrose:difftop}
for details.

We will here consider only the vacuum case, i.e. the case when $\ame$ is
Ricci flat, 
\begin{equation}\label{eq:vac}
\aR_{ab} = 0. 
\end{equation}
Let $M \subset \aM$ be a space--like hypersurface, i.e. a 
hypersurface with time--like normal $T$. We let $e_i$ be a frame on $M$ and
use indices $i,j,k$ for the frame components. 
Let $g,k$ be the induced metric and second fundamental form of 
$M \subset \aM$, where $k_{ij} =  \la \anabla_i e_j , T\ra$. 
Let $t$ be a time function on a neighborhood of $M$. Then we can introduce
local coordinates $(t,x^i, i=1,2,3)$ on $\aM$ so that $x^i$ are coordinates
on the level sets $M_t$ of $t$. This defines the coordinate vector field
$\partial_t$ of $t$. Alternatively we can let $M_t = \bfi(t,M)$ where 
$\bfi: \Re \times M \to \aM$ is a 
1-parameter family of imbeddings of an abstract 
3--manifold $M$. Then $\partial_t = \bfi_{*} d/dt$ where $d/dt$ is the 
coordinate derivative on $\Re$. 

Define the lapse $N$ and shift $X$
w.r.t. $t$
by $\partial_t = NT + X$. Assume that $T$ is future oriented so that $N >
0$. A 3+1 split of equation (\ref{eq:vac}) gives the Einstein vacuum 
constraint equations
\begin{subequations}\label{eq:constraint}
\begin{align}
R - |k|^2 + (\tr k)^2 &= 0 , \label{eq:constraint-ham} \\
\nabla_i \tr k - \nabla^j k_{ij}  &= 0 , \label{eq:constraint-mom}
\end{align}
\end{subequations}
and the Einstein vacuum evolution equations
\begin{subequations}\label{eq:evolution}
\begin{align}
\Lie_{\partial_t} g_{ij} &= -2 \Lapse k_{ij} + \Lie_{\Shift} g_{ij} , 
\label{eq:evolution-g}\\
\Lie_{\partial_t} k_{ij} &= - \nabla_i \nabla_j \Lapse + \Lapse(R_{ij} 
+ \tr k
k_{ij} - 2 k_{im}k^m_{\ j} ) + \Lie_{\Shift} k_{ij}  . \label{eq:evolution-k}
\end{align}
\end{subequations}
where $\Lie_{\partial_t}$ denotes Lie--derivative\footnote{for a tensor
$b$ on $M$, with $b_{ij} = b(e_i, e_j)$, we have $\Lie_{\partial_t} b_{ij} = \partial_t (b_{ij}) - b([\partial_t,
e_i],e_j) - b(e_i, [\partial_t, e_j])$} w.r.t. $\partial_t$. In
case $[\partial_t , e_i]=0$, $\Lie_{\partial_t}$ can be replaced by
$\partial_t$. 

A triple $(M,g,k)$ consisting of a 3--manifold $M$, a
Riemannian metric $g$ on $M$ and a symmetric covariant 2--tensor $k$ 
is a {\bf \index{vacuum data set}{vacuum data set}} for the Einstein equations if it solves
(\ref{eq:constraint}). 

\begin{definition}
Let $(M,g,k)$ be a vacuum data set. 
\begin{enumerate}
\item A  vacuum spacetime $(\aM, \ame)$ is called a {\bf \index{vacuum 
extension}{vacuum extension}} of $(M,g,k)$ if there there is an imbedding $\bfi$
with time--like normal $T$
of $(M,g, k)$ into $(\aM, \ame)$ so that 
$g = \bfi^* \ame$ and $k = - \bfi^* (\anabla T)$.
\item
A globally hyperbolic vacuum spacetime
$(\aM, \ame)$ is called a 
{\bf vacuum \index{Cauchy development}{Cauchy development}} of 
$(M, g, k)$ if there is an imbedding $\bfi$
with time--like normal $T$
of $(M,g, k)$ into $(\aM, \ame)$ so that $\bfi(M)$ is a Cauchy surface in 
$(\aM, \ame)$, $g = \bfi^* \ame$ and 
$k = - \bfi^* (\anabla T)$. If $(\aM, \ame)$ is {\em maximal} in the class of 
vacuum Cauchy 
developments of $(M, g , k)$ then $(\aM, \ame)$ is called the
{\bf maximal vacuum Cauchy development} (MVCD) of $(M, g, k)$. 
In the following, when convenient, we will identify $M$ with $\bfi(M)$. 
\end{enumerate}
\end{definition}
The Einstein vacuum 
equations are not hyperbolic in any standard sense due to the coordinate
invariance (``general covariance'') of the equation $\aR_{ab} = 0$. 
Nevertheless, the Cauchy problem
for the Einstein vacuum equation is well posed in the following sense. 

\begin{thm}[Choquet--Bruhat and Geroch \cite{ch-b:geroch:glob}]
\label{thm:ChBG}
Let $(M,g,k)$ be a vacuum data set. Then there is a unique, up to
isometry,  maximal vacuum Cauchy development (MVCD) 
of $(M, g, k)$. If $\phi: M \to 
M$ is a diffeomorphism, the MVCD of $(M, \phi^* g, \phi^* k)$ is isometric
to the MVCD of $(M,g,k)$. 
\end{thm}

The proof relies on the fact that in spacetime harmonic coordinates,
$\square_{\ame} x^{\alpha} = 0$, the Ricci tensor is of the form
\begin{equation}\label{eq:Ricciharm}
\aR_{\alpha\beta}^{(h)} = - \half \square_{\ame} \ame_{\alpha\beta} +
S_{\alpha\beta}[\ame, \partial \ame], 
\end{equation}
where $\square_{\ame}$ is the scalar wave operator in $(\aM, \ame)$. 
Hence the Einstein vacuum equations in spacetime harmonic
coordinates is a quasi--linear hyperbolic system and therefore the Cauchy
problem\footnote{Note that the  Einstein equations in spacetime harmonic
gauge should be viewed as an evolution equation for $(g,k,N,X)$.}
for $\aR_{\alpha\beta}^{(h)} = 0$ is well posed and 
standard results give local existence. One proves that
if the constraints and gauge
conditions are satisfied initially, they are preserved by the evolution.
This together with
a Zorn's lemma argument gives the existence of a MVCD. Uniqueness is proved
using the field equations to get a contradiction to the Haussdorff property,
given a pair of non--isometric vacuum Cauchy developments,
which are both maximal w.r.t. the
natural partial ordering on the class of Cauchy developments. 

A spacetime $(\aM, \ame)$ is said to satisfy the {\bf time--like 
convergence condition} (or {\bf \index{strong energy condition}{strong energy condition}}) if 
\begin{equation}\label{eq:TCC}
\aR_{ab} V^a V^b \geq 0, \quad  \text{ for all $V$ with } \ame_{ab} V^a V^b
\leq 0 .
\end{equation}
Globally hyperbolic spacetimes with compact Cauchy surface and satisfying
the time--like convergence condition are often called ``cosmological
spacetimes'' in the literature, following \cite{bartnik:cosmological}.
Here we will use the term spatially compact to refer to the existence of
a compact Cauchy surface. 
A spacelike hypersurface $(M,g)$ in $(\aM, \ame)$ has 
constant mean curvature if $\nabla_i \tr_g k = 0$, cf.  subsection \ref{sec:CMC} below.

We end this subsection with by stating a few facts 
about Killing fields.
\mnote{check this statement, is CMC necessary}
\begin{prop}[\protect{\cite{fischer:etal:linstab}}] Let $M$ be a compact
manifold and let $(M,g,k)$ be a constant mean curvature vacuum
data set on $M$ with MVCD $(\aM, \ame)$. Let $Y$ be a Killing
field on $(\aM, \ame)$ and let 
$Y = Y_{\perp} T + Y_{\parallel}$ be the splitting of $Y$ into its
perpendicular and tangential parts at $M$. 
Then 
$(Y_{\perp}, Y_{\parallel})$ satisfy the conditions 
\begin{enumerate}
\item \label{point:nonflat}
$Y_{\perp}  = 0$, $\Lie_{Y_{\parallel}} g = 0$, 
$\Lie_{Y_{\parallel}} k =0$, in case $g$ is non--flat or $k \ne 0$.
\item \label{point:flat}
$Y_{\perp}$ is constant and $\Lie_{Y_{\parallel}} g = 0$ if $g$ is
flat and $k=0$. 
\end{enumerate}
On the other hand, given $Y_{\perp}, Y_{\parallel}$ on $M$ satisfying 
conditions
\ref{point:nonflat}, \ref{point:flat} 
above, there is a unique Killing field $Y$ on $\aM$, 
with $Y= Y_{\perp}T + Y_{\parallel}$ on $M$.
\qed
\end{prop}

\begin{prop} Let $(\aM, \ame)$ be a globally hyperbolic 
spacetime.
\begin{enumerate}
\item \label{point:cmc}
Assume that $(\aM, \ame)$ satisfies the time--like convergence
condition and contains a compact Cauchy 
surface $M$ with constant
mean curvature. Then either $(\aM, \ame)$ is a metric product $M\times \Re$ or
any Killing field $Y$ on $(\aM, \ame)$ is tangent to $M$. In particular, if
$(\aM, \ame)$ is vacuum and has a nonzero Killing not tangent to $M$, then
$(\aM, \ame)$ is flat. 
\item \label{point:cpt}
Assume a compact group $G$ acts by isometries on $(\aM, \ame)$. Then the
action of $G$ is generated by space--like Killing fields and $(\aM,
\ame)$ is foliated by Cauchy surfaces invariant under the action of $G$. 
\item \label{point:strict}
\mnote{implicit in \cite{berger:etal:T2}?}
Assume that $(\aM, \ame)$ is 3+1 dimensional. 
Let $M$ be a Cauchy surface in $\aM$, let 
$Y$ be a Killing field on $\aM$ and assume $Y$ is
strictly spacelike, $\ame(Y,Y)>0$,  on  $M$. 
Then $Y$ is strictly spacelike on
$\aM$. 
\end{enumerate}
\end{prop}
\begin{proof} 
Point \ref{point:cmc} is a well known consequence of the uniqueness result
for constant mean curvature hypersurfaces of Brill and Flaherty
\cite{brill:flaherty}, cf. \cite{marsden:tipler}.
Point \ref{point:cpt} is essentially 
\cite[Lemma
1.1]{berger:etal:T2}. For the proof, note that 
as $G$ is compact we can construct
a $G$ invariant time function on $\aM$ by averaging any global time function 
$t$ 
on $\aM$ w.r.t. the $G$ action, cf. the proof of \cite[Lemma
1.1]{berger:etal:T2}. The level sets of the averaged time function are Cauchy
surfaces and are invariant under the action of $G$. The result follows. 

The following argument for point \ref{point:strict} is due to Alan Rendall\footnote{private
communication, 1999}.
Let $\mathcal N = \{p \in \aM : \ame(Y,Y) =0\}$ and assume for a
contradiction $\mathcal N$ is nonempty. Choose $p \in \mathcal N$ and a time
function $t$ on $\aM$ so that $t(M)=0$ and $t(p) > 0$. Let $A$ denote the
intersection of the past of $p$ with the future of $M$ and let 
$t_1 = \inf \{ t(q): q \in A \cap \mathcal N\}$. The set $\mathcal N$ is
closed and by global hyperbolicity $A$
is compact and hence $t_1 >0$ and 
there is a $q \in A \cap \mathcal N$ with $t(q)=t_1$. 
If $Y(q)$ is nonzero and null, then using the equation $Y^a\nabla_a (Y^b Y_b)
= 0$ which holds since $Y$ is Killing,
gives a null curve 
of points in $\mathcal
N$ where $Y$ is null. Following this into the past, shows that there is a $q
\in \mathcal N$ with $t(q) < t_1$, which gives a contradiction. 
In case $Y(q) = 0$, the linearization of $Y$ acts by isometries on
$T_q\aM$, and as the sphere of null directions in $T_q\aM$
is two dimensional it leaves a
null direction fixed. Using the exponential map shows that the action of $Y$
near $q$ leaves a null geodesic invariant, along which $Y$ must be zero or 
null. This leads to a contradiction as above. 
\end{proof}

\subsection{Cosmic Censorship}\label{sec:SCC}
Theorem \ref{thm:ChBG} proves uniqueness of the MVCD of a given data set
$(M, g, k)$. However, examples show that the MVCD may fail to 
be maximal in the class of all vacuum extensions, i.e. there exist examples
of vacuum data sets $(M,g, k)$ with vacuum extensions $(\aM, \ame)$ 
such that the MVCD of $(M,g,k)$ is a strict subset of $(\aM, \ame)$. 
\begin{example}\label{ex:misner}
Consider the $n+1$--dimensional 
Minkowski space $\Re^{n,1}$ with metric $\eta = - dt^2 + (dx^1)^2 + \cdots
+(dx^n)^2$, let $I^+(\{0\})$ be 
the interior of the future light cone. $I^+(\{0\})$ is globally hyperbolic
with the hyperboloids as Cauchy surfaces, and with the mantle of the 
light cone as Cauchy horizon. Let $\Gamma$ be a cocompact discrete
subgroup of the Lorentz group $\SO(n,1)$. Then the quotient
space $\aM = \Gamma \backslash I^+(\{0\})$ is a globally hyperbolic,
spatially compact spacetime. By choosing $\rho$ to be the Lorentzian distance
from the origin, we get 
$\ame = - d\rho^2 + \rho^2\gamma$ where $\gamma$ is the standard hyperbolic metric on
the compact quotient $M = \Gamma \backslash \bfH^n$. $(\aM, \ame)$ is the MVCD
of the vacuum data set $(M,\gamma, -\gamma)$. 

In case $n=1$, $\bfH^1 = \Re$, and a fundamental domain for $\Gamma$
can be found which intersects the null boundary of $I^+(\{0\})$ in an open
interval. Therefore if $n=1$, there is a nontrivial extension of $\aM$,
which is still flat, but which fails to be globally hyperbolic, cf. figure
\ref{fig:misner}. This spacetime is known as the Misner universe.
The maximal extension is unique in this case.
\begin{figure}
\centering
\includegraphics[height=2in]{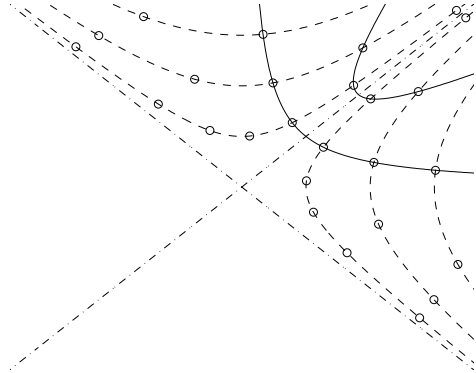}
\caption{The 1+1 dimensional Misner universe, showing a few orbits of
$\SO(1,1)$ (dashed lines) and orbits of 
a discrete subgroup $\Gamma \subset \SO(1,1)$ 
(circles). The boundary of a
fundamental domain for $\Gamma$ is plotted with solid lines.}
\label{fig:misner}
\end{figure}

If $n> 1$, the ergodicity of the geodesic flow on $(M, \gamma)$ can be shown 
to prevent the existence of a spacetime extending $(\aM, \ame)$
\cite{ishibashi:hyperbolic}. 

Higher dimensional examples of flat globally hyperbolic spatially compact
spacetimes which admit nontrivial non--globally hyperbolic extensions can be 
constructed by taking products of the $n=1$ Misner universe 
with the flat torus. 
\end{example}

A maximal vacuum extension may be non--unique, as is shown by the 
Taub-NUT example, cf. \cite{chrusciel:isenberg:nonisom}. 
If the MVCD of a vacuum data set is not a maximal vacuum extension, any
extension of it must fail to satisfy the intuitively reasonable causality
requirement of global hyperbolicity. 

According to physical intuition, causality violations should be rare. This
leads to the idea of cosmic censorship, essentially due to Penrose, see
\cite{penrose:unsolved} for discussion. One
way of stating this, 
relevant to the class of spacetimes we are concerned with
here, is the following form of the strong cosmic censorship
conjecture. 
\begin{conj}[Strong Cosmic Censorship]\index{cosmic censorship}{} 
\label{conj:SCC}
Let $M$ be a compact manifold of dimension $3$. 
Then for {\em generic} vacuum data sets
$(M,g,k)$, the maximal vacuum Cauchy development of $(M,g,k)$
is equal to the maximal vacuum
extension of $(M,g, k)$. 
\end{conj}
In the case of asymptotically flat spacetimes (describing 
isolated systems in general relativity), 
the so--called weak cosmic censorship 
conjecture states
that naked singularities (i.e. singularities which can be seen by an 
observer at infinity) should not occur generically, see the review paper by
Wald \cite{wald:WCCsurvey} for a discussion of the status of the weak cosmic 
censorship conjecture. The work of  Christodoulou, see
\cite{christodoulou:instnaked} and references therein, see also the discussion 
in \cite[\S 5]{wald:WCCsurvey}, establishes weak cosmic
censorship in the class of spherically symmetric Einstein--scalar field 
spacetimes, but also gives examples of initial data such that the Cauchy
development has a naked singularity. 
\mnote{mention Dafermos here}
For earlier surveys on the strong cosmic censorship
conjecture, see \cite{isenberg:SCC-survey} and \cite{chrusciel:SCC-survey}.

The Penrose inequality, giving a lower bound on the ADM mass in terms of 
the area of a horizon in black hole spacetimes, 
was derived by a heuristic argument
assuming the validity of the weak cosmic censorship
conjecture. 
The proof of the Riemannian version of the Penrose inequality by
Huisken and Ilmanen
\cite{huisken:ilmanen:penrose-announce,huisken:ilmanen:penrose},  
gives indirect support for the conjecture. 

Let $(\aM, \ame)$ be a spacetime and let $M \subset \aM$ be a 
space--like hypersurface.
The Cauchy horizon $H(M)$ is the boundary of
the domain of dependence $D(M)$, 
cf. Appendix \ref{sec:causality} for definition. 
If $M$ is compact without boundary, then every point of $H^+(M)$ lies on a past
inextendible null geodesic and every point of $H^-(M)$ lies on a future
inextendible null geodesic, where $H^+(M)$ and $H^-(M)$ are the future and
past components of $H(M)$, respectively.

Let $(\aM, \ame)$ be a maximal vacuum extension of a vacuum data set
$(M,g,k)$ with $M$ compact, and let $D(M) \subset \aM$ be the MVCD of
$(M,g,k)$. If $D(M) \ne \aM$, then the Cauchy horizon $H(M)$ is nonempty. 
One approach to SCC is to study the
geometry of Cauchy horizons in vacuum spacetimes and to prove rigidity
theorems as a consequence of extendibility of $D(M)$. 

Isenberg and Moncrief proved for analytic vacuum or electrovac
spacetimes, with analytic Cauchy
horizon $H(M)$, that under the additional assumption 
that $H(M)$ is ruled by closed null geodesics, there is a nontrivial
Killing field which extends to $D(M)$, see 
\cite{isenberg:moncrief:exceptional,isenberg:moncrief:symmetries}.
This result was generalized to the $C^{\infty}$ case by Friedrich
et. al. \cite{friedrich:etal:cauchy-rigid}. 
As spacetimes with Killing fields are non--generic, this may be viewed as
supporting evidence for the SCC. 

In the class of Bianchi spacetimes (i.e. spatially locally homogenous
spacetimes, cf. section \ref{sec:bianchi}), it has
been proved by Chru\'sciel and Rendall \cite{chrusciel:rendall:SCC},
generalizing work by Siklos \cite{siklos:whimper} in the analytic case, 
that any $C^{\infty}$ Bianchi spacetime which
contains a compact locally homogenous 
Cauchy horizon is a Taub spacetime, 
cf. section
\ref{sec:bianchi}, (\ref{eq:Taub-cond}) for definition. 
This result may be viewed as
a version of SCC in the class of Bianchi spacetimes. 
In this context, it is worth mentioning that 
work by Chru\'sciel and Galloway \cite{chrusciel:galloway:nondiff}
gives examples which indicate that Cauchy horizons may be
non--differentiable,
generically.

\begin{conj}[Bartnik \protect{\cite[Conjecture 2]{bartnik:cosmological}}]
\label{conj:bartnik} 
Let $(\aM, \ame)$ be a spatially compact globally hyperbolic 
spacetime satisfying the 
time--like convergence condition (\ref{eq:TCC}). Then if $(\aM, \ame)$ is
time--like geodesically complete, $(\aM, \ame)$ splits isometrically as a
product $(\Re \times M, - dt^2 + g)$. 
\end{conj}
If the Bartnik conjecture \ref{conj:bartnik} is true, then any vacuum, globally hyperbolic, 
spatially compact spacetime, is either flat and covered by $\Re \times T^3$
or is has an inextendible time--like geodesic which ends after a finite proper
time, i.e. it is time--like geodesically incomplete. 
A sequence of points approaching the ``end'' of a
finite length inextendible geodesic is often thought of as approaching a
singularity.
See \cite{galloway:banach} for a discussion of the status of the
Bartnik conjecture. 

Inextendibility of $D(M)$ can be detected by monitoring the asymptotic
behavior of curvature invariants 
such as the Kretschmann scalar $\kappa$, 
defined by $\kappa = \aR_{abcd} \aR^{abcd}$.
If $\kappa$ blows up along causal geodesics, then
$D(M)$ fails to be extendible, and therefore proving blowup for $\kappa$ for 
generic spacetimes is an approach to proving SCC. 
This is the method used in the proof of cosmic
censorship for the class of polarized Gowdy spacetimes 
\cite{chrusciel:SCC:gowdy},  cf. section \ref{sec:gowdy},
and is likely to be important also in the cases with less
symmetry. 
The structure of the horizon and extensions in the polarized Gowdy 
class can be very complicated as shown by the work of Chru\'sciel et. al., see 
\cite{chrusciel:SCC-survey} for discussion, see also
\cite{chrusciel:isenberg:nonisom}. 
It was proved by Ringstr\"om, cf. Theorem \ref{thm:mixmaster}, 
that for vacuum
Bianchi spacetimes of class A, either the spacetime is Taub, cf. section 
\ref{sec:bianchi}, or $\kappa$
blows up at the singularity.

\subsection{The evolution equations}\label{sec:evoleqs}
A solution to the vacuum \index{Einstein evolution equations}{Einstein evolution equations} 
with initial data is a curve $t \mapsto (g(t), k(t), N(t), X(t))$ 
defined on some interval $(T_0,T_1)$, satisfying (\ref{eq:evolution}). 

Every regular solution $(g,k,N,X)$ to the vacuum evolution equations
(\ref{eq:evolution})
with initial data solving the vacuum constraint equations, 
gives a vacuum spacetime.  This is due to the fact that the 
constraint quantities
\mnote{check whether we should have $D_i = \nabla_i \tr k - \nabla^j
k_{ij}$?} 
\begin{align*}
B &= R + (\tr k)^2 - |k|^2 \\
D_i &= \nabla_i \tr k - \nabla^j k_{ij} 
\end{align*}
evolve according to 
a symmetric hyperbolic system 
and energy estimates together with an
application of the Gronwall inequality allow one to show the the constraints 
are satisfied during the time of existence of the solution curve. Now
the fact that the Einstein vacuum equation is equivalent to the system of
constraint and evolution equations 
shows that the spacetime $(\aM, \ame)$
constructed from the curve $(g,k,N,X)$ by letting 
$\aM = (T_0, T_1) \times M$, and setting 
$$
\ame = - N^2 dt^2 + g_{ij}(dx^i + X^i dt) (dx^j + X^j dt)
$$
is a solution to the Einstein vacuum equations (\ref{eq:vac}). 
 
Note that in order for the
solution to be well defined, it is necessary to specify the lapse and shift
$(N,X)$, 
either as functions on spacetime $M \times (T_0 ,T_1)$ or as functions of the
data, $N= N[g,k], X = X[g,k]$; this may be viewed as a gauge
fixing for the Einstein equations. 

The choice of lapse and shift is crucial for
the behavior of the solution curve. In particular, a foliation constructed for 
a particular choice of $N,X$ may develop singularities which are not caused
by any singular or irregular nature of the Cauchy development. Consider 
for example the Gauss foliation condition $N=1, X=0$. Then the hypersurface 
$M$ flows in the direction of its unit normal and $M_t$ is simply the level 
set of the Lorentzian distance function $t(p) = d(M,p)$. The 
foliation $\{M_t\}$ will
develop singularities precisely at the focal set of $M$, which in general
will be nonempty, even in Minkowski space.

Many authors have considered hyperbolic reformulations of the 
Einstein equations, see the paper by Friedrich \cite{friedrich:hyperbolic} 
for discussion, see also 
\cite{frittelli:reula:hyperbolic,iriondo:etal:ashtekarhyp,iriondo:etal:ashtekardyn}
for related work. 
The development of singularities for hyperbolic systems presents a serious
obstacle to the numerical treatment of the Einstein evolution equations
using hyperbolic reformulations, 
see \cite{alcubierre:shock,alcubierre:masso}
for discussion and examples.  It is therefore necessary to consider also
gauges which make (\ref{eq:evolution}) into an elliptic--hyperbolic system. 

\subsection{Constant mean curvature foliations}\label{sec:CMC}
\index{constant mean curvature}{}
A particularly interesting choice of gauge condition for the lapse function
is given by the constant mean curvature (CMC) \index{CMC}{} condition 
$$
\nabla_i \tr k = 0 , 
$$ 
i.e. the level sets $M_t$  of the time function $t$ are assumed to be
hypersurfaces of constant mean curvature in $(\aM, \ame)$. 
If $(\aM, \ame)$ is globally hyperbolic, spatially compact and 
satisfies the time--like
convergence condition (\ref{eq:TCC}), 
then for $\tau \in \Re$, either there is at
most one Cauchy surface with mean curvature  $\tr k = \tau$ or 
$(\aM, \ame)$ splits
as a product, cf. \cite{brill:flaherty}. 

This indicates that the mean curvature $\tr k$ 
may be useful as a time function on
$(\aM, \ame)$ in the spatially compact case. 
Setting $t = \tr k$ leads, using (\ref{eq:constraint}--\ref{eq:evolution})
to the lapse equation 
\begin{equation}\label{eq:lapse}
- \Delta N + |k|^2 N = 1 .
\end{equation}
Local wellposedness for the Einstein evolution equations in CMC and more
general elliptic time gauges given by a lapse equation of the form 
$-\Delta N + |k|^2 N = NT(h)$
for given spacetime functions $h$ has been proved by Choquet-Bruhat and York 
\cite{ChB-York:ellhyp}. We may call this gauge the prescribed mean curvature (PMC)
gauge. The proof is based on writing a wave equation for the
second fundamental form, a technique which was also used in the work of
Choquet-Bruhat and Ruggeri \cite{ChB:ruggeri} on harmonic time gauge.  
The \index{prescribed mean curvature}{prescribed mean curvature} condition,
has been used in the $\U(1)\times \U(1)$ symmetric case by Henkel \cite{henkel:PMC} who proved
global existence in this gauge. An important advantage of using PMC gauge is
that one avoids the problem of proving existence of a CMC Cauchy surface.

Equation (\ref{eq:lapse}) supplemented by an elliptic shift gauge makes the
Einstein evolution equations (\ref{eq:evolution}) into an
elliptic--hyperbolic
system of evolution equations. Let $\hme$ be a given smooth metric on
$M$, and let 
$V^i = g^{mn} (\Gamma^i_{mn} - \hGamma^i_{mn})$. Then $-V^i$ is the tension
field of the identity map $\Id: (M,g) \to (M, \hme)$, and the CMC condition
coupled with the spatially \index{harmonic coordinate condition}{harmonic
coordinate condition} 
$$
V^i = 0
$$
gives a new elliptic gauge condition for the Einstein equations, 
\index{CMCSH gauge}{CMCSH gauge} 
\cite{andersson:moncrief:local}. In \cite{andersson:moncrief:local}, it is
shown that the Einstein equations in CMCSH gauge forms an 
\index{elliptic-hyperbolic system}{elliptic-hyperbolic system} 
for the Einstein equations, which is well posed
in $H^s \times H^{s-1}$, $s > n/2+1$. 

The maximal slicing condition $\tr k = 0$ is of interest mainly for the
asymptotically flat case. This was used in the proof of the nonlinear
stability of Minkowski space by Christodoulou and Klainerman
\cite{christo:klain:stability}, cf. the discussion in section \ref{sec:3+1}.
Due to the ``collapse of the lapse'' phenomenon, see \cite{beig:omurchadha},
the maximal foliation is not expected to cover the whole MVCD except in the
small data case. See \cite{marsden:tipler} for a discussion of maximal
slices. Asymptotically flat spacetimes satisfying certain restrictions on
the causal structure are known to contain maximal hypersurfaces
\cite{bartnik:maximal:1984}. The existence of CMC hypersurfaces in
asymptotically flat spacetimes was considered in \cite{andersson:iriondo}. 

The mean curvature operator satisfies a geometric maximum principle, see
\cite{andersson:etal:maxprin} for a proof of this under weak regularity. 
This allows one to use barriers to prove existence of constant mean curvature
hypersurfaces. A spacetime is said to have {\bf crushing singularities} if
there are sequences of Cauchy surfaces with mean curvature $\tr k$ 
tending uniformly to $\pm \infty$.
Gerhardt \cite{gerhardt:CMC} proved, using a barrier argument,
that any spacetime satisfying
(\ref{eq:TCC}) with crushing
singularities is globally foliated by CMC hypersurfaces. 
These facts indicate that the CMC foliation condition is an interesting
time gauge for the Einstein evolution equation. 
See also
\cite{gerhardt:pmc,ecker:huisken:prescCMC,bartnik:acta,bartnik:cosmological,henkel:PMC}
for results relevant to existence of CMC hypersurfaces.

Let $R[g]$ be the scalar curvature. 
A 3--manifold $M$ is said to be of Yamabe type $-1$ if it admits no metric
with $R[g] = 0$ (and hence no metric with nonnegative scalar curvature), 
of Yamabe type $0$ if it
admits a metric with $R[g] = 0$ but no metric with $R[g] = 1$ and of Yamabe
type $+1$ if it admits a metric with $R[g] = 1$,
cf. \cite[Definition 9]{fischer:moncrief:hamred}. 

If the Cauchy surface $M$ is of Yamabe type $-1$, 
it follows from the constraint equation that $(\aM, \ame)$ 
cannot contain a maximal (i.e. $\tr k = 0$)
Cauchy surface and therefore one expects that (if the dominant energy
condition holds) the maximal time
interval of existence for (\ref{eq:evolution}) in CMC time
is of the form (after a time orientation)  $(-\infty, 0)$ with $\tau \nearrow
0$ corresponding to infinite expansion. 

If $M$ is of Yamabe type $0$,  then one expects that either 
the maximal CMC time interval is $(-\infty, 0)$ (possibly after a change of
time orientation) or $(\aM, \ame)$ splits as a
product, and therefore in the vacuum case is covered by $\Re \times T^3$ 
with the flat metric. Finally in case $M$ is of Yamabe type $+1$, one expects the
maximal CMC time interval to be $(-\infty, \infty)$, i.e. the spacetime evolves
from a ``big bang'' to a ``big crunch''. 
This is formalized in the ``closed
universe recollapse conjecture'' of Barrow, Galloway and Tipler
\cite{barrow:etal:recollapse}. 

\begin{conj}[Constant mean curvature foliations]\label{conj:CMC}
Let $M$ be a compact
3--manifold and let $(M,g,k)$ be a vacuum
data set on $M$, with constant mean curvature. 
The Cauchy problem for the Einstein vacuum evolution
equations with data $(M,g,k)$ 
has global existence in the constant mean curvature time gauge, 
i.e. there is a CMC
foliation in the MVCD $(\aM,\ame)$ of $(M,g,k)$, containing $M$, 
with mean curvature taking all values in
$(-\infty,\infty)$ in case $M$ has Yamabe type $+1$ and in case 
$M$ has Yamabe type $0$ or $-1$, taking all values in
$(-\infty,0)$ (possibly after a change of time orientation).
\end{conj}
\begin{remark}\label{rem:globrem} Conjecture \ref{conj:CMC} has been stated in
essentially this form 
by Rendall \cite[Conjecture
1]{rendall:cosmologicalCMC}, see also Eardley and Moncrief
\cite[Conjecture C2]{eardley:moncrief:CC}
for a closely related statement.

Note that as $\tau \searrow -\infty$, the
past focal distance of the (unique) CMC surface with mean curvature $\tau$
tends to zero, and
hence the foliation exhausts the past of $M$. It follows that in case $M$ has
Yamabe type $+1$, then if Conjecture \ref{conj:CMC} is true, $(\aM, \ame)$ is
globally foliated by CMC hypersurfaces. In case $M$ has Yamabe type $0$ or
$-1$ on the other hand, there is the possibility that the CMC foliation does
not cover all of $\aM$, due to the fact that as the mean curvature $\tau
\nearrow 0$, the CMC hypersurfaces are expected to avoid black holes, by
analogy with the behavior of CMC and maximal 
hypersurfaces in the Schwarzschild spacetime. 
See \cite{rendall:cosmologicalCMC} for further remarks and conjectures
related to this.
\end{remark}

If one were able to prove Conjecture \ref{conj:CMC}, then as remarked in
\cite{eardley:moncrief:CC}, this would give the possibility of attacking the 
Cosmic Censorship Conjecture using PDE methods. There are no known
counterexamples to Conjecture \ref{conj:CMC} for vacuum spacetimes.
However, Isenberg and  Rendall \cite{isenberg:rendall:noCMC} 
give an example of dust spacetimes, 
not covered by a CMC foliation.
Bartnik
\cite{bartnik:cosmological} gave an example of a spatially compact, 
globally hyperbolic 
spacetime satisfying the time--like convergence condition,  
which
contains no CMC Cauchy surface.
It is an interesting open question whether or not similar counter examples
are possible in the vacuum case. The gluing results for solutions of the
constraint equations, see
\cite{isenberg:etal:glue,isenberg:etal:glueII,corvino:glue} may lead to the
construction of such examples. 

The CMC conjecture \ref{conj:CMC} has been proved in a number of cases for
spacetimes with symmetry, in the
sense of the existence of a group $G$ of isometries acting (locally) on 
$(\aM, \ame)$ by isometries and with space--like orbits. 
In the case of Bianchi IX,  cf. section
\ref{sec:bianchi}, which has Yamabe type $+1$, the closed universe recollapse
conjecture and consequently Conjecture
\ref{conj:CMC}, was proved
by Lin and Wald \cite{lin:wald:recollapse}.
In the case of 2+1--dimensional vacuum spacetimes with cosmological
constant, the conclusion of Conjecture \ref{conj:CMC} is valid
\cite{andersson:etal:2+1grav}. In the 2+1 case, 
the CMC foliations are global.  

The work of Rendall 
and Burnett, see \cite{burnett:rendall:maximal} and references therein, 
proves under certain restrictions on the matter that a maximal, globally
hyperbolic, spherically symmetric spacetime,
\mnote{does existence of one CMC follow from work of Henkel in this case?}
which contains a CMC Cauchy surface diffeomorphic to
$S^2 \times S^1$, is globally foliated by CMC hypersurfaces with mean
curvature taking on all real values. 

We end this subsection by mentioning the \index{harmonic time gauge}{harmonic
time gauge}
condition, $\square_{\ame} t = 0$ or 
$$
\ame^{ab} \aGamma_{ab}^0 = 0
$$
In case $X = 0$, this is equivalent to the condition 
$N = \sqrt{\det g}/\sqrt{\det e}$ where
$e_{ij}$ is some fixed Riemannian metric on $M$. The Einstein evolution
equations with $X = 0$ were proved to be hyperbolic with this time gauge by
Choquet--Bruhat and Ruggeri \cite{ChB:ruggeri}.  This time gauge also
appears in the work of Gowdy and is used in the analysis of the Gowdy
spacetimes as well as in the numerical work of Berger, Moncrief et. al. on
Gowdy and $\U(1)$ spacetimes, cf. sections \ref{sec:gowdy} and
\ref{sec:U(1)}. 

The papers of Smarr
and York \cite{smarr:york,smarr:york:radiation} contain an interesting
discussion of gauge conditions for the Einstein equations. See also section 
\ref{sec:3+1} for
a discussion of the spatial harmonic coordinate gauge and the
survey of Klainerman and Nicolo \cite{klainerman:nicolo:review} for further
comments on gauges. 

\subsection{The Einstein equations as a system of quasi--linear
PDE's}\label{sec:PDE}
As mentioned above, the Einstein vacuum equations in spacetime harmonic
coordinates form a quasi--linear hyperbolic system of the form 
\begin{equation}\label{eq:QL}
-\half 
\square_{\ame} \ame_{\alpha\beta} + S_{\alpha\beta} [\ame, \partial \ame] =
0 .
\end{equation}
The system (\ref{eq:QL}) is a quasi--linear wave equation, 
quadratic in the first order derivatives 
$\partial \ame$ and with top order symbol depending only on the field $\ame$
itself. Standard results show that the Cauchy problem is \index{well
posed}{well posed} in
Sobolev spaces $H^s \times H^{s-1}$, $s> n/2 +1$. This was first proved for
the Einstein equations by Hughes, Kato and Marsden
\cite{hughes:kato:marsden}. It is also possible to prove this for
elliptic--hyperbolic systems formed by the Einstein evolution equations
together with the CMC--spatial harmonic coordinates gauge, see
\cite{andersson:moncrief:local}, see also section \ref{sec:3+1}  for a
discussion of the spatial harmonic coordinates gauge. 
\mnote{check if this is done in sec \ref{sec:3+1}}

Recent work using harmonic analysis methods by Bahouri and Chemin
\cite{bahouri:chemin}, Klainerman and Rodnianski
\cite{KR:ricci,KR:rough,KR:causal} and Smith and Tataru \cite{smith:tataru:NLW},
see also \cite{tataru:ICM2002}
has pushed
the regularity needed for systems of the above type on $\Re^{n,1}$
to $s > (n+1)/2$ for $n \geq 3$. 
In order to get well posedness for $s$ lower than the values given above, it
is likely one needs to exploit some form of the null condition, 
\cite{sogge:survey,sogge:lectures,tataru:ICM2002}. 
The null condition for equations of the form 
$\square_{\eta} u = F[u, \partial u]$ on Minkowski space 
states roughly that the symbol 
of the non--linearity $F$ cancels null vectors. For a discussion of the null
condition on a curved space background, see
\cite{sogge:nullcurved,smith:sogge}. Counter examples to well posedness for
quasi--linear wave equations with low regularity data are given by Lindblad, see
\cite{lindblad:quasi-counter}. 

The standard example of an equation which satisfies the null condition is 
$\square u = \partial_{\alpha} u \partial_{\beta} u \eta^{\alpha\beta}$
where $\square$ is the wave operator w.r.t. the Minkowski metric $\eta$ on
$\Re^{n,1}$. 
This equation is well posed for data in $H^s \times H^{s-1}$
with $s > n/2$ \cite{klainerman:selberg:optimal}, 
and has global existence for small data for $n \geq 3$. On the 
other hand, the equation $\square u = (\partial_t u)^2$ which does not
satisfy the null condition can be shown to have a finite time of existence for
small data in the 3+1 dimensional case.

For quasi--linear wave equations which satisfy an appropriate form of the
\index{null condition}{null condition} \cite{hormander:book},
global existence for small data is known in 3+1 dimensions. 
The Einstein equations, however,
are not known to satisfy the null condition in any 
gauge. In particular, it can be seen that in spacetime harmonic coordinates, 
the Einstein equations do not satisfy the null condition. 
However, the
analysis by Blanchet and Damour \cite{blanchet:damour:hereditary} of the
expansion of solutions of Einstein equations in perturbation series around
Minkowski space indicates that the logarithmic terms in the gravitational
field in spacetime harmonic coordinates, arising from the violation of the 
null condition, can be removed after a (nonlocal) gauge transformation to
radiative coordinates where the coordinate change 
depends on the history of the field. A similar analysis can be done for the 
Yang--Mills (YM) equation in Lorentz gauge. 
It may further be argued that the small data, global existence proof of
Christodoulou and Klainerman for the Einstein equations exploits properties
of the Einstein equations related to the null condition. In a recent paper, 
Lindblad and Rodnianski \cite{lindblad:rodnianski:weaknull}
have introduced a weak form of the null condition and
shown that the Einstein equations satisfy this. 

Global existence is known for several of the classical field equations such
as certain nonlinear Klein--Gordon (NLKG) equations 
and the YM equation  on $\Re^{3,1}$ (proved by Eardley and Moncrief,
\cite{eardley:moncrief:YMI,eardley:moncrief:YMII}). The proofs for NLKG and
the proof of Eardley and Moncrief for YM use light cone estimates
to get apriori $L^{\infty}$ bounds. The proof of Eardley and Moncrief used
the special properties of YM in the radial gauge. 
This method was also used in the global
existence proof for YM on 3+1--dimensional,
globally hyperbolic spacetimes by
Chru\'sciel and Shatah \cite{chrusciel:shatah:YM}. Klainerman and Machedon
\cite{klainerman:machedon:YM} were able to prove that the YM equations on
$\Re^{3,1}$ in
Coloumb gauge satisfy a form of the null condition and 
are well posed in energy space $H^1 \times L^2$. They were then able to 
use the
fact that the energy is conserved to prove global existence for
YM. See also \cite{klainerman:survey} for an overview of these
ideas and some related conjectures. 

An important open problem for the classical field equations is
the global existence problem for the \index{wave map equation}{wave map
equation}  
(nonlinear $\sigma$--model, hyperbolic harmonic map equation). 
This is an equation for a map $\Re^{n,1} \to N$, where $N$ is some complete
Riemannian manifold, 
\begin{equation}\label{eq:wavemap}
\square_{\eta} u^A + \Gamma_{BC}^A (u) \partial_{\alpha} u^B \partial_{\beta}
u^C \eta^{\alpha\beta} = 0 .
\end{equation}
Here $\Gamma$ is the Christoffel symbol on $N$. 

The wave map equation satisfies the null condition 
and hence we have small data global existence
for $n \geq 3$. Small data global existence is also known for
$n=2$. 
Further, scaling arguments provide counterexamples to global
existence for $n \geq 3$, whereas $n=2$ is critical with respect to
scaling.
For $n \geq 2$, global existence for ``large data'' is known only
for symmetric solutions, and in particular, the global existence problem for
the wave map equation (\ref{eq:wavemap}) is open for the case $n=2$. 
For the case $n=1$, global
existence can be proved using energy estimates or light cone estimates.  
Tao proved \cite{tao:2+1wavemap} that the wave map equation in 2+1 dimensions
with the sphere as target is globally well--posed for small energy. The proof
is based on a new local well--posedness in $H^1\times L^2$ 
and conservation of energy. This result has been generalized to the
case of hyperbolic target by Krieger \cite{krieger:2+1:wavemap}. 
Tataru \cite{tataru:roughmap} has proved that the wave map equation in $n$+1
dimensions is locally well--posed in $\dot
H^{n/2} \times \dot H^{n/2-1}$ for general target imbedded in $\Re^m$. (Here
$\dot H^s$ is the homogenous Sobolev space. See 
\cite{tataru:roughmap} for the precise definition of well--posedness used).
See also \cite{shatah:survey,tataru:wavemap} for surveys. \mnote{see also Tao
notes on dispersive equations} \mnote{mention work of Struwe on wave maps
w. symmetry} 

The above discussion shows that the situation for the wave map equation is 
reminiscent of that for the Einstein equations,
cf. sections \ref{sec:gowdy}, \ref{sec:U(1)}. In particular, it is
interesting to note that equations of the wave map type show up in the
reduced vacuum Einstein equations for the Gowdy and $\U(1)$ problems.

\section{Bianchi}\label{sec:bianchi}
Let $(\aM, \ame)$ be a 3+1 dimensional spacetime with 3--dimensional local
isometry group $G$. 
Assume the action of $G$ is generated by space--like Killing
fields and that the orbits of $G$ in the universal cover of $\aM$ are
$3$--dimensional.  
This means there is a global foliation of $\aM$ by space--like
Cauchy surfaces $M$ with locally homogeneous induced geometry. 
Such spacetimes are known as \index{Bianchi spacetimes}{Bianchi spacetimes}.
The assumption of local homogeneity of the 3--dimensional Cauchy surfaces
means that a classification of the universal cover 
is given by the classification of 3--dimensional Lie algebras.

Let $e_a$, $a= 0,\dots,3$ be an ON frame on $\aM$, with $e_0 = u$,
a unit time--like normal to the locally homogeneous Cauchy surfaces, let
$\gamma_{ab}^c$ be the commutators of the frame,
$[e_a, e_b] =
\gamma_{ab}^c e_c$. Let the indices $i,j,k,l$ run over $1,2,3$. We may
without loss of generality assume that $[e_a, \xi_i] = 0$ where
$\{\xi_i\}_{i=1}^3$ is a basis for the Lie algebra $\frakg$ of $G$. 

Choose a time function $t$ so that $t_{,a} u^a =1$, i.e. the level sets of
$t$ coincide with the group orbits.
Restricting to a level set $M$ of $t$, 
the spatial part of the
commutators $\gamma_{ij}^k$ are the structure constants of $\frakg$.
These can be decomposed into a constant symmetric matrix $n^{kl}$ and a vector
$a_i$, 
$$
\gamma_{ij}^k = \eps_{ijl} n^{kl} + a_i \delta^k_j - a_j \delta^k_i .
$$ 
We will briefly describe the classification used in the physics literature,
cf. \cite{ellis:maccallum}, \cite[\S 1.5.1]{ellis:wainwright:book}.

The Jacobi identity implies $n^{ij} a_j = 0$ and by 
choosing the frame $\{e_i\}$ to diagonalize $n^{ij}$ and so that $e_1$ is
proportional to the vector $a_i$, 
we get 
\begin{equation}\label{eq:ndiag}
n^{ij} = \diag(n_1 , n_2 , n_3), \qquad  a_i = (\mathbf a,0,0) .  
\end{equation}
The 3--dimensional Lie algebras are divided into two classes by the
condition $\mathbf a = 0$ (class A) and $\mathbf a \ne 0$ (class B). The classes $A$ and
$B$ correspond in mathematical terminology 
to the unimodular and non--unimodular Lie algebras. 
If $n_2 n_3 \ne 0$, let the scalar $h$ be defined by 
\begin{equation}\label{eq:hdef}
\mathbf a^2 = h n_2 n_3 . 
\end{equation}
Table \ref{tab:bianchi-type} gives the classification of 
Bianchi geometries.
Note that the invariance of the Bianchi types under
permutations and sign changes of the frame elements has been used to 
simplify the presentation. 
\begin{table}
\caption{Bianchi geometries}
\begin{tabular}{lccc}
Class A & 
\begin{tabular}{lccc}
 Type & $n_1$ & $n_2$ & $n_3$ \\
\hline 
 I      & 0 & 0 & 0 \\
 II     & 0 & 0 & + \\
 VI$_0$ & 0 & - & + \\
 VII$_0$ & 0 & + & + \\
 VIII & - & + & + \\
 IX & + & + & + 
\end{tabular} & 
\ \ Class B & 
\begin{tabular}{lccc} 
Type & $n_1$ & $n_2$ & $n_3$ \\
\hline
 V   & 0 & 0 & 0 \\
 IV  &  0 & 0 & + \\
 VI$_h$&  0 & - & + \\
 VII$_h$& 0 & + & + 
\end{tabular} 
\end{tabular}
\label{tab:bianchi-type}
\end{table}
Here the notation VI$_0$, VII$_0$, VI$_h$, VII$_h$ refers to the value of
$h$ defined by (\ref{eq:hdef}).
In the list of Bianchi types I--IX, the missing type III is the same as
VI$_{-1}$. 

Due to the local homogeneity of the Cauchy surfaces $M$ in a Bianchi
spacetime, the topologies of the 
spatially compact Bianchi spacetimes can be classified using the
classification of compact manifolds
admitting Thurston geometries. 

The eight \index{Thurston geometries}{Thurston geometries} 
$S^3$, $E^3$, $H^3$,
$S^2 \times \Re$, $H^2 \times \Re$, Nil, $\tSL(2,\Re)$, Sol, 
are the maximal geometric structures on
compact 3--manifolds, see
\cite{scott:geometries,thurston:book} for background. Each compact
3--manifold with a Bianchi (minimal) 
geometry also admits a Thurston (maximal) geometry, and
this leads to a classification of the topological types of compact 
3--manifolds with
Bianchi geometry, 
i.e. compact manifolds of the form $X/\Gamma$ where $X$ is a
complete, simply connected 3--manifold with a Bianchi geometry and $\Gamma$
is a cocompact subgroup of the isometry group of $X$. It is important to
note that $\Gamma$ is not always a subgroup of the 3--dimensional Bianchi
group $G$. 

The relation between the Bianchi types admitting a compact quotient and the
Thurston geometries 
is given by Table \ref{tab:thurston-bianchi}.
For each Bianchi type we give only the maximal
Thurston geometries corresponding to it,
see \cite{fujiwara:etal:bianchi,kodama:canonical,koike:etal:compact-homog}
for further details and references. We make the following remarks
\begin{remark}\label{rem:bianchi-thurston}
\begin{enumerate}\renewcommand{\theenumi}{\roman{enumi}}
\item 
Let $(M,g)$ be a 3--dimensional space form with sectional
curvature $\kappa = -1, 0 , +1$. A spacetime $(\aM, \ame)$ with $\aM = M
\times (a,b)$ and a warped product metric $\ame = - dt^2 + w^2(t) g$, 
satisfying the perfect fluid Einstein
equations is called a (local) Friedmann--Robertson--Walker (FRW)
spacetime. Specifying the equation of state for the matter in the Einstein
equations leads to an ODE for $w$.
The FRW spacetimes play a central role in the standard
model of cosmology. In the vacuum case, only $\kappa = -1,0$ are possible, 
and in this case, 
the spatially compact 
local FRW spacetimes are for $\kappa = 0$, the flat spacetimes covered by  
$T^3\times \bfE^1$, a special case of Bianchi I, 
and for $\kappa = -1$, 
the local FRW spacetimes discussed
in example \ref{ex:misner}, which are Bianchi V. 
\item 
The Thurston geometry $S^2\times \Re$ 
admits no 3--dimensional group of isometries, and hence 
it does not
correspond to a Bianchi geometry, but to a Kantowski--Sachs geometry
\cite{kantowski:sachs} with symmetry group $\SO(3)\times \Re$. 
The other
cases with 4-dimensional symmetry group are vacuum LRS Bianchi 
I, II, III, VIII and IX, see \cite[p. 133]{ellis:maccallum}. 
\mnote{classification $G_4$ according to Email from Woei-Chet Lim}
\item \label{point:BianchiIII}
The type of geometry depends on the subgroup $\Gamma$ of the 
isometry group used to construct
the compactification. The isometry group in turn depends on the Bianchi
data. Note that $\tSL(2,\Re)$ is both maximal and minimal, whereas the
isometry group of $\bfH^2 \times \Re$ has dimension 4 and is therefore not a
minimal (Bianchi) group. 
\item \label{point:BianchiV} 
The compactifications of Bianchi V and VII$_h$, $h \ne 0$ are both
of the type discussed in Example \ref{ex:misner}, 
thus no anisotropy is allowed in the
compactification of Bianchi type V and VII$_h$, $h \ne 0$. 
\item 
The compactification of a Bianchi geometry, introduces new (moduli)
degrees of freedom, in addition to the dynamical degrees of
freedom, see \cite{kodama:canonical} and references therein for
discussion, see also \cite{chrusciel:completeness}. 
The resulting picture is complicated and does not appear to have
been given a definite treatment in the literature. 
\end{enumerate}
\end{remark}

\begin{table}
\caption{Bianchi and Thurston geometries}
\begin{tabular}{cccp{2in}}
Bianchi type  & class & Thurston geometry & comments			\\
\hline
I		& A	&$\bfE^3$		&		\\
\hline
II		& A	&  Nil			&		\\
\hline
III = VI$_{-1}$	& B	&\begin{tabular}{c}\  \\ $\bfH^2 \times \bfE^1$\\
				$\tSL(2,\Re)$ \end{tabular} 
& cf. Remark \ref{rem:bianchi-thurston}  \ref{point:BianchiIII}. 
\\	
\hline
IV		& ---	& ---			& no compact quotient \\
\hline
V		& B	& $\bfH^3$		& 
cf. Remark
\ref{rem:bianchi-thurston} \ref{point:BianchiV}
\\
\hline
VI$_0$		& A	& Sol			&			\\
\hline
VI$_h$, $h\ne 0,-1$& --- & --- 			& no compact quotient	\\
\hline
VII$_0$		&  A	&$\bfE^3$		&			\\
\hline
VII$_h$, $h\ne 0$& B	&$\bfH^3$		& 
cf. Remark
\ref{rem:bianchi-thurston} \ref{point:BianchiV} \\
\hline
VIII		& A	& $\tSL(2,\Re)$		&			\\
\hline
IX		& A	& $S^3$			&			\\
\hline
\end{tabular}
\label{tab:thurston-bianchi}
\end{table}

In the rest of this section, we will concentrate on
class A Bianchi spacetimes. We will also refrain from considering the
moduli degrees of freedom introduced by the compactification, as it can be
argued that these are not dynamical. 

Let 
the expansion tensor $\theta_{ij}$ be given by 
$$
\theta_{ij} = \nabla_j u_i  , 
$$
(i.e. $\theta_{ij} =  - k_{ij}$ 
where $k_{ij}$ is the second fundamental form).
Decompose $\theta_{ij}$ as 
$
\theta_{ij} = \sigma_{ij} + H \delta_{ij} 
$
where $H = \theta/3$, $\theta = g^{ij} \theta_{ij}/3$. 
Then $H$ is the Hubble scalar.
We are assuming (\ref{eq:ndiag}) and in the vacuum class A case it follows
that $\sigma_{ij}$ is diagonal \cite[p. 41]{ellis:wainwright:book}.

Under these assumptions, $\gamma_{ab}^c = \gamma_{ab}^c(t)$ and we may
describe the geometry of  $(\aM, \ame)$ completely in terms of
$\gamma_{ab}^c$ or equivalently in terms of the 3--dimensional commutators 
$\gamma_{ij}^k$ and the expansion tensor given in terms of $\sigma_{ij},
\theta$. 

By a suitable choice of frame, $\sigma_{ij}$ and $n_{ij}$ can be assumed
diagonal. Since $\sigma_{ij}$ is traceless and diagonal, it can be described
in terms of the variables 
$$
\sigma_+ = \half ( \sigma_{22} + \sigma_{33}), \quad \sigma_- =
\frac{1}{2\sqrt{3}} (\sigma_{22} - \sigma_{33} ) 
$$
and similarly, $n_{ij}$ can be represented by 
$$
(n_1, n_2, n_3 ) = \frac{1}{2\sqrt{3}}(n_{11}, n_{22}, n_{33} ) .
$$
The curvature $b_{ij} = Ric(e_i, e_j )$ can be written in terms of $n_{ij}$, 
$b_{ij} = 2 n_i^{\ k} n_{kj} - n_k^{\ k} n_{ij}$ and is therefore diagonal in
the chosen frame. Decomposing $b_{ij}$ into trace--free and trace parts
$s_{ij}, k$ we find that $s_{ij}$ can be represented by two variables $s_+, s_-$.
 
Next we introduce the dimensionless variables (following Hsu and Wainwright
\cite{hsu:wainwright:1989}, but normalizing with $H$ instead of $\theta$),
$$
(\Sigma_+, \Sigma_-, N_1, N_2 , N_3 ) = (\sigma_+, \sigma_-, n_1, n_2, n_3)/H
$$
Similarly, we set 
$$
(S_+, S_-) = (s_+, s_-) /H^2
$$
In addition to these choices we also define a new time $\tau$ by $e^{\tau}
= \ell$, $\ell$ the length scale factor, or $dt/d\tau = 1/H$. 

Clearly the equations are invariant under permutations 
$(\Sigma_i) \to P(\Sigma_i)$, $(N_i) \to P(N_i)$. 
Cyclic permutations of $(N_i), (\Sigma_i)$ correspond to rotations through
$2\pi/3$ in the $\Sigma_+, \Sigma_-$ plane. 

We now specialize to the vacuum case. Then (for $H \ne 0$)
the Einstein equations are
equivalent to the following system of ODE's for
$\Sigma_+$, $\Sigma_-$, $N_1$, $N_2$, $N_3$ (where the $'$ denotes
derivative w.r.t. the time coordinate $\tau$): 
\begin{subequations}\label{eq:evol}
\begin{align}
N_1' &= (q - 4 \Sigma_+) N_1 , \\
N_2' &= (q +2 \Sigma_+ + 2 \sqrt{3} \Sigma_-) N_2 , \\
N_3' &= (q +2 \Sigma_+ - 2 \sqrt{3} \Sigma_-) N_3 , \\
\Sigma_+' &= - (2-q) \Sigma_+ - 3 S_+ , \\
\Sigma_-' &= - (2-q) \Sigma_- - 3 S_- , 
\end{align}
\end{subequations}
where 
\begin{align*}
q &= 2 ( \Sigma_+^2 + \Sigma_-^2 ) , \\
S_+ &= \frac{2}{3} [ (N_2 - N_3)^2 - N_1 ( 2N_1 - N_2 - N_3) ] , \\
S_- &= \frac{2}{\sqrt{3}} (N_3 - N_2) ( N_1 - N_2 - N_3) .
\end{align*}
The Hamiltonian constraint (\ref{eq:constraint-ham}) 
is in terms of these variables 
\begin{equation}\label{eq:hamconstr}
\Sigma_+^2 + \Sigma_-^2 + 
[ N_1^2 + N_2^2 + N_3^2 - 2(N_1 N_2 + N_2 N_3 + N_3 N_1) ] = 1 .
\end{equation}

With our conventions, 
if $(\tau_-, \tau_+)$ is the maximal time
interval of existence for the solution to (\ref{eq:evol}), then $\tau \to
\tau_-$ corresponds to the direction of a singularity. For all non--flat
vacuum Bianchi models except Bianchi IX, the spacetime undergoes an infinite
expansion as  $\tau \to\tau_+$ and is geodesically complete in the expanding
direction, cf. \cite[Theorem 2.1]{rendall:globhom} which covers the vacuum
case as a special case. For Bianchi IX, on the
other hand, $\tau \to \tau_+$ corresponds to $H
\to 0$, as follows from the proof of the closed universe recollapse
conjecture for Bianchi IX by Lin and Wald \cite{lin:wald:recollapse}, 
and hence to the dimensionless variables becoming ill defined. 

We will now review the basic facts for Bianchi types I, II and IX. 
\bigskip

\noindent{\bf I:} Kasner. The Hamiltonian (\ref{eq:hamconstr}) 
constraint reads 
\begin{equation}\label{eq:hamI}
\Sigma_+^2 + \Sigma_-^2 = 1 , 
\end{equation}
and the induced metric on each time slice is flat. These spacetimes can
be given metrics of the form 
\begin{equation}\label{eq:kasner}
ds^2 = - dt^2 + \sum_i t^{2p_i} dx^i \tens dx^i  ,
\end{equation}
where $\sum_i p_i = 1$, $\sum_i p_i^2 = 1$, the \index{Kasner
relations}{Kasner relations}, which
correspond to the equation (\ref{eq:hamI}). The $\Sigma_{\pm}$ and $p_i$ are
related by $\Sigma_+ = \frac{3}{2} (p_2 + p_3) - 1$, $\Sigma_- =
\frac{\sqrt{3}}{2}(p_2 - p_3)$. Clearly, the \index{Kasner circle}{Kasner
circle}
given by (\ref{eq:hamI}) consists of fixed points to the system
(\ref{eq:evol}). The points $T_1, T_2, T_3$, with coordinates 
$(-1,0)$, $(1/2 , \pm \sqrt{3}/2)$ correspond to flat spacetimes of type I 
or VII$_0$ (quotients of Minkowski space). 
\bigskip

\begin{figure}
\centering
\includegraphics[width=\textwidth-1in,height=2in]{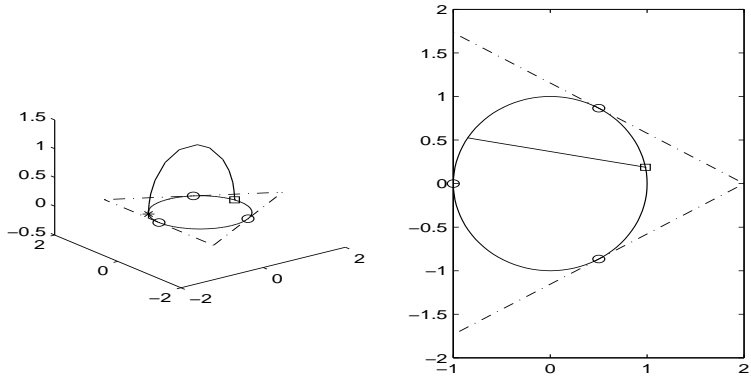}
\caption{A type II solution, the Kasner circle and the triangle for the 
Kasner billiard in the $(\Sigma_+,\Sigma_-)$--plane are shown.}
\label{fig:typeII}
\end{figure}

\noindent{\bf II:} By using the permutation symmetry
of the equations, we may assume $N_2 = N_3 =
0$. The solution curve which is a subset of the cylinder $\Sigma_+^2 +
\Sigma_-^2 < 1$ has a past endpoint on the longer arc connecting $T_2$ and $T_3$
and future endpoint on the shorter arc connecting these points see
figure \ref{fig:typeII}. This curve
realizes the so called Kasner map, cf. 
\cite[\S 6.4.1]{ellis:wainwright:book}. We define the Bianchi II variety as
the union of the three spheres in $\Re^5$ given by the solutions to the constraint
(\ref{eq:hamconstr}) with $(N_2,N_3) = 0$ and permutations thereof. The
Kasner circle is the intersection of the spheres.
\bigskip 

\noindent{\bf IX:} \index{Mixmaster}{Mixmaster}, 
characterized by $N_i > 0, i=1,2,3$. 
The heuristic picture is the following. The projection in the
$(\Sigma_+,\Sigma_-)$--plane of a generic orbit in the direction
$\tau \searrow \tau_-$ moves into the Kasner circle and stays there,
undergoing an infinite sequence of bounces, which are approximately given
by the \index{Kasner billiard}{Kasner billiard}, 
cf. figure \ref{fig:billiard}. 
This picture is supported by numerical studies of the full Bianchi IX
system, see eg. \cite{berger:etal:algorithm}. 

The Kasner billiard
is the dynamical system given by mapping a (non--flat) 
point $p$ on the Kasner circle to the point on the Kasner circle which is the
end point of the type II orbit starting at $p$. This map can be
described as follows. Let $B$ be the nearest corner to $p$
of the triangle shown
in figure \ref{fig:billiard}. The ray starting at $B$ through $p$
intersects the Kasner circle in a point $q$, which is the image of $p$
under the Kasner map, see also \cite[Fig. 6.13]{ellis:wainwright:book}. 
Iterating this construction gives a sequence of points $\{ p_i\}$ on the
Kasner circle, which we may call the Kasner billiard. 

\begin{figure}
\centering
\includegraphics[width=\textwidth]{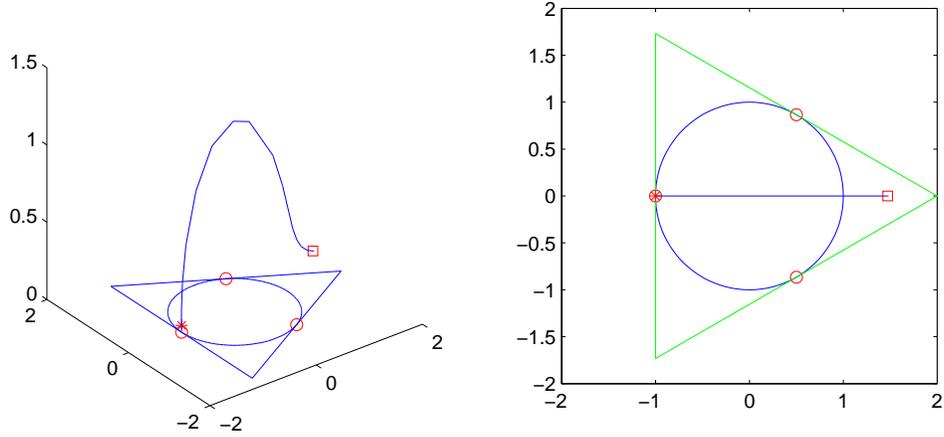}
\caption{A Taub type IX solution, the Kasner circle and the
triangle for the Kasner billiard in the $(\Sigma_+, \Sigma_-)$--plane 
are shown.}
\label{fig:taubNUT}
\end{figure}

\begin{figure}
\centering
\includegraphics[width=\textwidth-1.2in,height=2in]{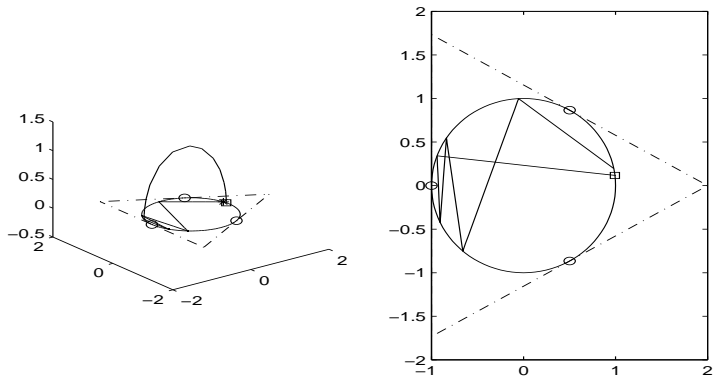}
\caption{Bianchi IX orbit showing a few bounces. The vertical axis is
$N_1$. The Kasner circle and the
triangle for the Kasner billiard in the $(\Sigma_+,\Sigma_-)$--plane
are shown.}
\label{fig:billiard}
\end{figure}

The exceptional orbits which do not exhibit this infinite sequence of
bounces are the Taub type IX solutions, cf. figure \ref{fig:taubNUT}. Up
to  a permutation these are given by the conditions 
\begin{equation}\label{eq:Taub-cond}
N_2 = N_3, \quad \Sigma_- = 0 .
\end{equation}
We call a Bianchi spacetime, 
satisfying
(\ref{eq:Taub-cond}) up to a permutation, a Taub spacetime. 

The past limit of the Taub type IX solution is the flat
point $(-1,0)$. The MVCD of Taub type IX data
has a smooth Cauchy horizon, and is extendible, the extension
being given by the so--called \index{Taub--NUT}{Taub-NUT} spacetimes. 
As shown by Chru\'sciel
and Rendall \cite[Theorem 1.2]{chrusciel:rendall:SCC}, 
these are the only Bianchi
IX spacetimes with a smooth Cauchy horizon, which gives a version of SCC
for this class. 
See \cite{chrusciel:rendall:SCC} for further details on the
status of SCC in the locally homogeneous case. 
Chru\'sciel and Isenberg proved that 
the MVCD of 
Taub type IX data has non--isometric maximal vacuum extensions, further
emphasizing its pathological nature, cf. \cite{chrusciel:isenberg:nonisom}.

From the point of view of the cosmic censorship conjecture, the following
theorem appears fundamental. A point which is a past limit point of
$(\Sigma_+(\tau),\Sigma_-(\tau),N_1(\tau),N_2(\tau),N_3(\tau))$ is called an
$\alpha$ limit point. We say that the approach to the singularity is
oscillatory, if the set of $\alpha$ limit points contains at least two points
on the Kasner circle, at least one of which is disctinct from the special 
points $T_1,T_2,T_3$, cf. \cite{rendall:mixmaster}. 
\begin{thm}[\cite{rendall:mixmaster,ringstrom:blowup}]
\label{thm:mixmaster} 
A vacuum Bianchi spacetime of class A
has exactly one of the properties
\begin{enumerate}
\item The Kretschmann scalar $\kappa = \aR_{\alpha\beta\gamma\delta}
\aR^{\alpha\beta\gamma\delta}$ satisfies 
$\limsup_{\tau \searrow \tau_-} |\kappa| = \infty$
\item The MVCD has a smooth Cauchy horizon
and the spacetime is a Taub spacetime.
\end{enumerate}
For non--Taub vacuum Bianchi VIII and IX spacetimes, 
the approach to the
singularity is oscillatory.
\end{thm} 

\begin{remark}
\begin{enumerate}\renewcommand{\theenumi}{\roman{enumi}}
\item Rendall \cite{rendall:mixmaster} proved the dichotomy in Theorem
\ref{thm:mixmaster} for all  Bianchi class A except VIII and IX. These cases
as well as the oscillatory behavior for type VIII and IX were 
proved by Ringstrom \cite{ringstrom:blowup}. Curvature blowup for
non--Taub Bianchi A models with perfect fluid matter, including stiff fluid
was proved by 
by Ringstrom \cite{ringstrom:attractor}.
\item 
\mnote{what about curvature blowup for Bianchi
  class B models???}
In class B it is only the exceptional model Bianchi VI$_{-1/9}$ which
  exhibits oscillatory behavior as has been argued by Hewitt et. al
  \cite{hewitt:etal:VI-1/9}. 
\item Theorem \ref{thm:mixmaster} shows that SCC holds in the class of vacuum
Bianchi class A spacetimes, also with respect to $C^2$ extensions.
\item See also Weaver \cite{weaver:VI0}
for a related result for Bianchi VI$_0$ with a magnetic field. Weaver proved
that the singularity in magnetic Bianchi VI$_0$ is oscillatory and that
curvature blows up as one approaches the singularity. It should be
noted that vacuum Bianchi VI$_0$, on the other hand, is non-oscillatory.
\end{enumerate}
\end{remark}

The dynamics of the Bianchi spacetimes has been studied for a long time,
from the point of view of dynamical systems. In particular, it is
believed that the Bianchi IX (mixmaster) solution is chaotic in
some appropriate sense, see
for example the paper by Hobill in \cite{ellis:wainwright:book} or the
collection \cite{hobill:book} as well as the work of Cornish and
Levin \cite{cornish:levin:farey},  for various points of view. However, it 
does not yet appear to be clear which is the appropriate definition of
chaos to be used, and no rigorous analysis exists for the 
full Bianchi IX system. In the course of the above mentioned work, 
approximations to the Bianchi dynamics have been
described and studied, such as the Kasner billiard (cf. fig. 
\ref{fig:billiard}, and the discussion above) and the BKL map, cf. 
\cite[\S 11.2.3]{ellis:wainwright:book}. 

\mnote{mention Misner and Chitre, Damour and Henneaux}

It has long been conjectured that the Bianchi II phase space is the
\index{attractor}{attractor}
for the Bianchi VIII and IX dynamics, see 
\cite{ellis:wainwright:book}. 
\mnote{find the correct ref to EW}
The phase space for the Bianchi II model is
given by the Hamiltonian constraint (\ref{eq:hamconstr}) together with 
the condition $N_1N_2 = N_2 N_3 = N_1 N_3 = 0$. This defines a variety 
consisting of the union of three 2--spheres in $\Re^5$. The 
Kasner circle is the intersection of these spheres. 
The conjecture has been proved in the Bianchi IX case. 
\begin{thm}[\protect{\cite{ringstrom:attractor}}]\label{thm:attractor}
The Bianchi II variety is the asymptotic attractor for vacuum Bianchi IX.
\mnote{did Ringstrom prove this for matter as well?} 
\end{thm}
This result goes a goes a long way towards proving the chaotic
nature of the Bianchi IX dynamics.

\begin{figure}[!tbp]
\begin{minipage}[t]{0.45\linewidth}
    \centering
    \includegraphics[width=2in]{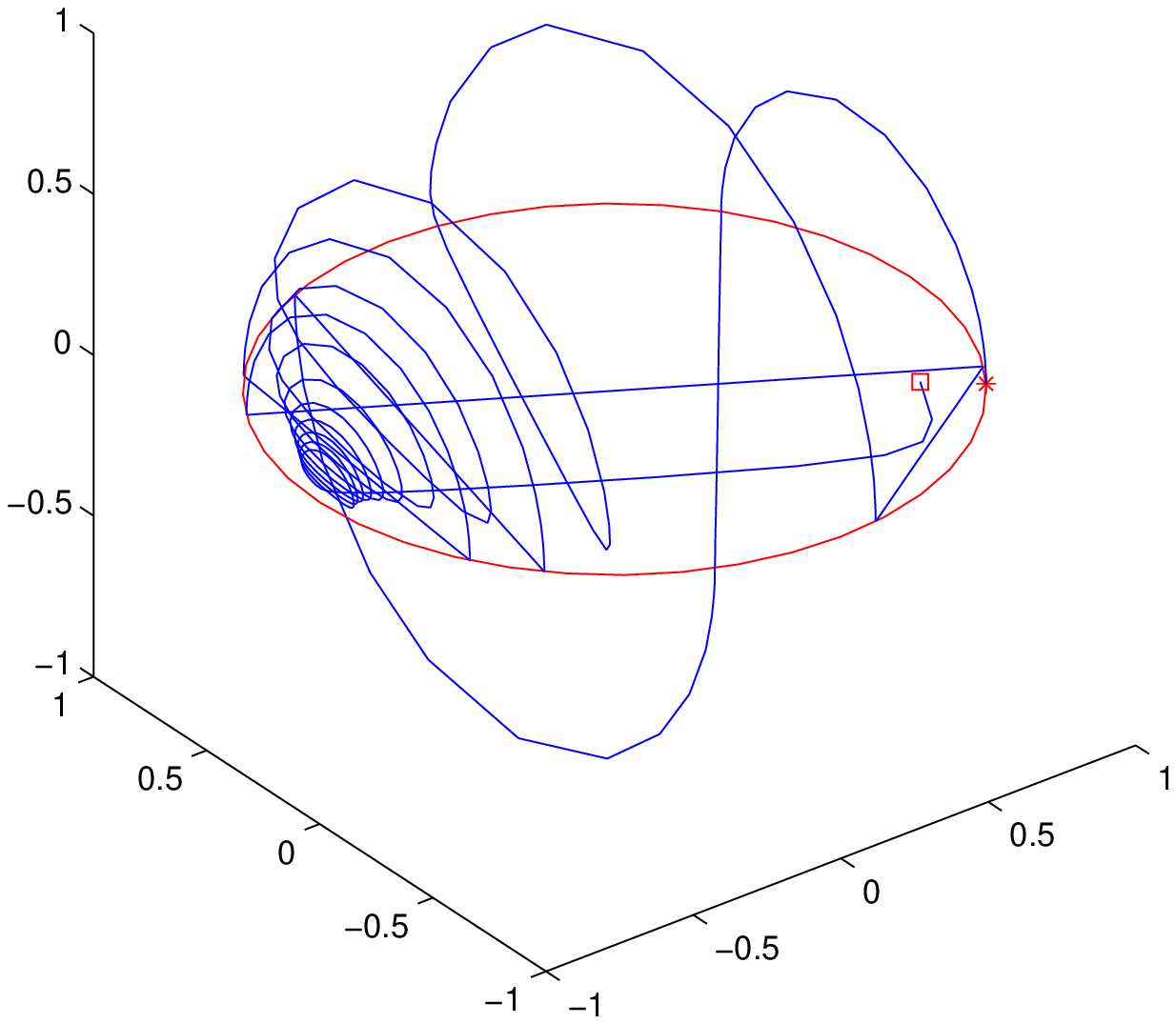}
\end{minipage}%
\begin{minipage}[t]{0.45\linewidth}
    \centering
\includegraphics[width=2in]{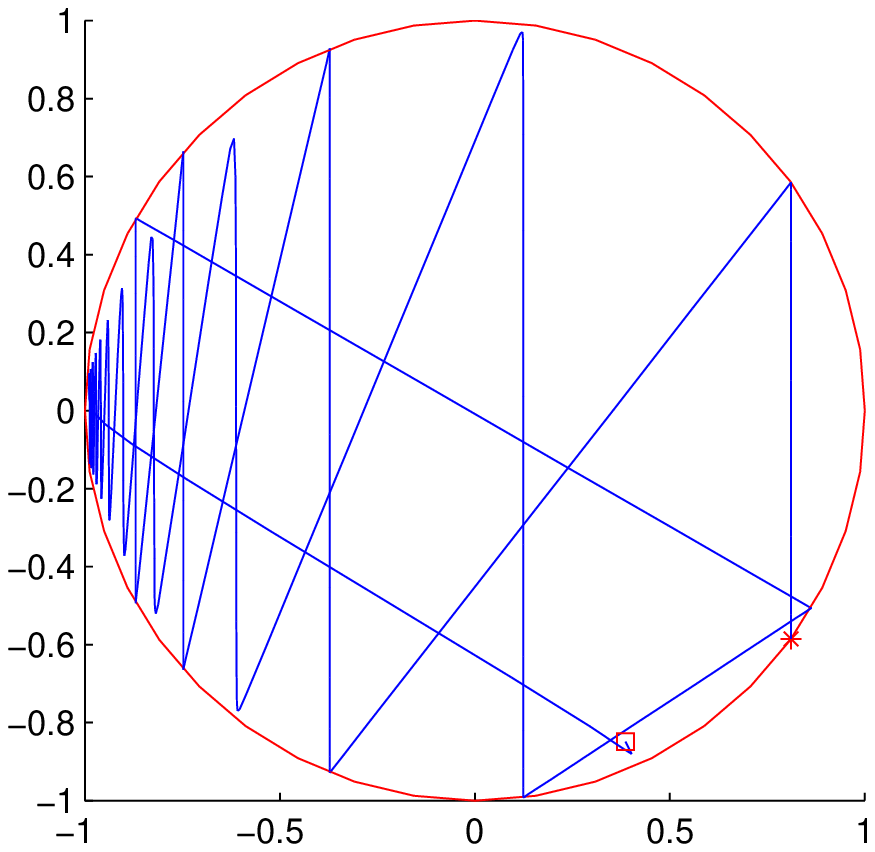}
\end{minipage}%
\label{fig:bianchiex}
\caption{An orbit of the Bianchi VI$_{-1/9}$ system, showing a few
bounces. Note that in contrast to the Bianchi IX system, the dynamics shows a 
combination of several types of bounces.}
\end{figure}

The questions of curvature blowup and oscillatory approach to the singularity
can be studied also in the case of Bianchi class B models. In this case we
have curvature blowup except for Bianchi III and V. 
\mnote{check this with JW}
However, in class B  
it is only the so--called exceptional Bianchi VI$_{-1/9}$ which, 
based on numerical work and qualitative 
analysis by Hewitt et. al \cite{hewitt:etal:VI-1/9} 
has an oscillatory singularity. 
\mnote{Hewitt et al \cite{hewitt:etal:VI-1/9} also have attractor conj. for
VI$_{-1/9}$} 
The exceptional Bianchi VI$_{-1/9}$ has the same number of degrees
of freedom as the most general class A models, Bianchi VIII and IX. 
Figure \ref{fig:bianchiex} shows an orbit of the Bianchi VI$_{-1/9}$ system. 

\mnote{mention Ringstrom future asympt for Bianchi VIII and other similar
results} 

\section{$G_2$}\label{sec:gowdy}

In this section we consider the case when $(\aM, \ame)$ is a
3+1--dimensional, spatially
compact, globally hyperbolic, vacuum spacetime, with a 2--dimensi\-onal
local isometry group $G_2$ with the action of $G_2$ generated by space--like
Killing fields.  
By passing to the
universal cover, we see that the non--degenerate orbits of the $G_2$--action are
2--dimensional homogeneous spaces and hence the induced metric on the orbits
must have constant curvature. The isometry group of the sphere $S^2$ has no
two dimensional subgroup and thus the orbits have geometry $\bfE^2$ or
$\bfH^2$. \index{$G_2$ spacetime}{}

The special case when the group $\U(1)\times \U(1)$ 
itself acts  on $\aM$ was considered by \index{Gowdy spacetime}{}
Gowdy \cite{gowdy:PRL,gowdy:1974}. 
In this case, it follows that orbits are compact (unless there is an extra
Killing field), and $M$ is covered by 
$T^3, S^3$ or $S^1\times S^2$. Suppose $p$ is a fix point for the action of
$\U(1)\times \U(1)$ on $M$, then $\U(1)\times \U(1)$ acts by isometries on
$T_pM$. 
As the isometry group $\SO(3)$ of $T_p M$ 
does not have a 2--dimensional subgroup, 
any degenerate orbit of $G$ must be a closed curve. 

Let $\aM$ be a bundle over
$S^1 \times \Re$ with compact 2--dimensional fiber $F$ and suppose that the
orbits of the $G$--action on the universal cover of $\aM$ cover the fibers 
$F$. If $F$ has geometry
$\bfE^2$ it follows that the Killing fields generating the $G$--action 
commute, and hence it is
natural, following Rendall \cite{rendall:gowdyCMC}, to use the term local
$\U(1)\times \U(1)$ symmetry for this situation. 

Space--times with local $\U(1)\times \U(1)$ symmetry have also been 
considered by Tanimoto
\cite{tanimoto:newgowdy}, 
who discussed the question of in which case a spacetime with local
$\U(1)\times \U(1)$ symmetry can be considered as a dehomogenization of a
Bianchi
spacetime. It can be seen from the structure of the Bianchi groups that the
only $\U(1)\times \U(1)$ symmetric spacetimes which have Bianchi (or
Kantowski--Sachs) limits are covered by $T^3$ or $S^3$, 
and in the case of $S^3$ (Bianchi IX), it is only
the Taub metrics that admit a $\U(1)\times \U(1)$ action by isometries. 
If we consider the case with local $\U(1)\times \U(1)$--symmetric case on the other hand, 
then all Bianchi
models except Bianchi VIII and XI, and in case of Bianchi VIII and IX the
Taub metrics, can be viewed as limits of locally
$\U(1)\times \U(1)$--symmetric models, and therefore these serve as
dehomogenization of the Bianchi models.

For simplicity, we concentrate in the rest of this section
only the case of $\U(1)\times \U(1)$ symmetric
spacetimes,  
and assume that the
twist constants vanish, i.e.  i.e. $\xi_1 \wedge
\xi_2 \wedge d\xi_1 = \xi_1 \wedge \xi_2 \wedge d\xi_2 = 0$, where 
$\xi_1,\xi_2$ are one forms dual to the generators of the $\U(1)\times \U(1)$ action. 
This is a nontrivial restriction only in case 
$M \cong T^3$, see \cite[p. 211]{gowdy:1974}. Such spacetimes are known in
the literature as Gowdy spacetimes. 
We further specialize to the case 
with Cauchy surface $M \cong T^3$ 

Let $\tau,x$ be coordinates on the 1+1 dimensional Lorentzian orbit
space $(\U(1)\times \U(1)) \backslash \aM$ and let $A(\tau,\theta)$ be the area of the orbit. 
Gowdy showed that in the non--twisted case, with $M \cong T^3$,  
there are no degenerate orbits, i.e. $A \ne 0$, and further, 
the level sets of $A$ in $(\U(1)\times \U(1)) \backslash \aM$ are space--like. 
We may therefore 
choose coordinates so that $A = 4 \pi^2 e^{-\tau}$, and choose the metric on the orbit as $Ah$ where $h = h(\tau,x)$ is a
unit determinant metric. 
$\aM$ is causally incomplete in the
direction $\tau \nearrow \infty$, which corresponds to a cosmological
singularity. 

By construction, the metric on the orbit, $h = h(\tau,x)$ 
is a unit determinant metric which is
constant on each orbit. It therefore represents an element of the
Teichm\"uller space $\Teich(T^2)$. The space $\Teich(T^2)$ with the
Weil--Peterson metric is isometric to the hyperbolic plane $\bfH^2$.
The identification of $\Teich(T^2)$ with $\bfH^2$ gives a map $u: 
(\U(1)\times \U(1))  \backslash \aM \to \bfH^2$. This can be realized concretely for example by
using the model for $\bfH^2$ with metric 
\begin{equation}\label{eq:hypmet}
dP^2 + e^{2P} dQ^2 
\end{equation}
and letting $u = (P,Q)$ with 
$$
P = \ln(h_{11}) , \qquad
Q = e^{-P} h_{12} .
$$
This is the parametrization that is used in the numerical work of Berger
and collaborators. 
%
Thus the Gowdy (as well as the general $\U(1)\times \U(1) $ symmetric) 
Einstein equations on
$M \cong T^3 \times \Re$ can be viewed as equations for the evolution of a
loop in $\bfH^2$. The velocity of a point $u(x,\tau)$ in $\bfH^2$ is given by  
$\vhyp(x,\tau) = \sqrt{\la
\partial_\tau u , \partial_\tau u \ra}$. 
Define the asymptotic velocity $\vhathyp(x)$
by $\vhathyp(x) = \lim_{\tau \to \infty} \vhyp(x,\tau)$, when the limit exists.
 
The spacetime metric may then be written in the form\footnote{We have assumed here that the
Killing fields are hypersurface orthogonal, see \cite{chrusciel:U1xU1:1990}
for the most general form of the $\U(1)\times \U(1)$ symmetric metric on $T^3
\times \Re$ with vanishing twist.}  
\begin{equation}\label{eq:gowdy-metric}
\begin{split}
\ell_0^{-2} ds^2 &=  e^{-\lambda/2+ \tau/2}  
(-e^{-2\tau} d\tau^2+dx^2 ) \\
 &\quad  +e^{-\tau } [e^Pdy_2^2+2e^P Qdy_2 dy_3 
+(e^PQ^2+e^{-P}) dy_3^2 ]
\end{split}
\end{equation}
where $\ell_0$ is the unit of physical length. 
Here $(\tau, \theta) \in \Re \times S^1$ are coordinates on the orbit space, 
$y^A, A = 2,3$ are coordinates on the orbit, $0 \leq y^A \leq 2
\pi$. The $G$ invariance implies that we can assume all metric components
depend on $\tau,\theta$ only. 
Let $e = dx^2 + (dy^2)^2 + (dy^3)^2$. 
The lapse function $N$ satisfies  
$N = \sqrt{\det(g)/\det(e)}$ and hence the time function $\tau$ is spacetime
harmonic, cf. subsection \ref{sec:CMC}.

Let $\eta = -d\tau^2 + e^{2\tau} dx^2$. The Einstein evolution 
equations take the form 
\begin{equation}\label{eq:gowdy-wavemap}
\eta^{\alpha\beta} ( \partial_{\alpha\beta} u^a + \Gamma_{bc}^a(u)
\partial_{\alpha} u^b \partial_{\beta} u^c ) = 0 ,
\end{equation}
where $\Gamma_{bc}^a$ are the Christoffel symbols on $\bfH^2$ with the
metric (\ref{eq:hypmet}). 
The system (\ref{eq:gowdy-wavemap}) is 
supplemented by a pair of equations for $\lambda$, which are implied by the 
Einstein constraint equations, and which are used to reconstruct the 3+1
metric $\ame$. 

Equation (\ref{eq:gowdy-wavemap}) is a semilinear hyperbolic system, which 
resembles the wave--map equation, (\ref{eq:wavemap}). 
Energy estimates
or light cone estimates prove global existence on $(0,\infty) \times S^1$. 
The wave operator $-\partial_\tau^2 + e^{-2\tau} \partial_x^2$ 
degenerates as $\tau \nearrow \infty$, which corresponds to a
singularity in the 3+1 spacetime $(\aM, \ame)$, since the area $e^{-\tau}$ 
of the orbit tends to zero.

The energy $E = \half \int_{S^1} 
\la \partial_\tau u , \partial_\tau u\ra + e^{-2\tau} \la \partial_x u , \partial_x u
\ra  = E_K + E_V$, where $E_K, E_V$ are the kinetic and potential energy
terms, respectively, satisfies 
$$
\partial_\tau E = - 2E_V .
$$
Therefore $E$ is monotone decreasing, with a rate determined by $E_V$. This
shows immediately that there is a sequence of times $(\tau_k)$, $\lim_{k\to
\infty} \tau_k = \infty$, so that $E_V(\tau_k) \to 0$ as $k \to \infty$, which
indicates that the scale--free variables 
$(e^{-\tau} \partial_x P, e^{-\tau} \partial_x Q)$ become insignificant for the
dynamics as $\tau \to \infty$. This leads to the idea that the Gowdy system
behaves asymptotically as a dynamical system in the likewise 
scale--free variables $(\partial_\tau P, \partial_\tau Q)$. 
This heuristic is, as we shall see, 
supported by numerical work and some rigorous results.  

We now briefly discuss some results and open problems for the Gowdy
and $\U(1) \times \U(1)$ symmetric spacetimes. 
The following result proves Conjecture \ref{conj:CMC} for this class. 
The first global existence result for Gowdy spacetimes with topology 
$T^3\times \Re$ was due to 
Moncrief \cite{moncrief:global:gowdy:1981}, who proved that vacuum Gowdy
spacetimes with the stated topology are globally foliated by level sets of
the area function, assuming the existence of a
compact level set of the area function. This assumption was removed by
Chrus\'ciel \cite{chrusciel:U1xU1:1990} who also studied global properties of
the area function for Gowdy spacetimes on $S^3$ and $S^2 \times S^1$. A class
of ``nongeneric'' metrics still remains to be studied, see
\cite{chrusciel:U1xU1:1990}.  
The first result concerning global CMC foliations in Gowdy spacetimes 
was proved by Isenberg and Moncrief 
\cite{isenberg:moncrief:CMCgowdy} for the case of vacuum Gowdy spacetimes.
Recently the following was proved. 
\begin{thm}[Andr\'easson, Rendall, Weaver \protect{\cite{ARW:T2CMC}}]
\label{thm:T2CMC}
Nonflat spatially compact spacetimes with $(\U(1)\times \U(1)) $
symmetry and Vlasov matter, 
are globally foliated by CMC hypersurfaces with mean curvature taking on all
values in $(-\infty, 0)$.
\end{thm}

It was conjectured by Belinskii, Khalatnikov and Lifschitz 
\cite{lifschitz:khalatnikov:1963} that in a generic spacetime with a
cosmological singularity, spatial points will decouple as one
approaches the singularity, and spatial derivatives become insignificant
asymptotically. This leads to the idea that asymptotically near the
singularity, the dynamics of
the gravitational field should be explained by a family of ODE systems. 
This is a very rough idea. Yet, as it turns out, this principle
appears to hold in the cases we have been able to study.  
\mnote{refer to other place where BKL is discussed}
The 
idea of BKL 
has been specialized and reformulated by among others Eardley, Liang and
Sachs \cite{eardley:liang:sachs} and Isenberg and Moncrief
\cite{isenberg:moncrief:asymptGowdy}, 
into the notion of asymptotically velocity term dominated (AVTD)
singularities. Roughly, an \index{AVTD}{AVTD} 
solution approaches asymptotically, 
at generic spatial points, 
the solution to an ODE, the
parameters of which depend on the spatial point. In particular, in an AVTD
spacetime, locally 
near a fixed spatial point the spacetime approaches a Kasner limit, with 
parameters depending on the spatial point. The asymptotic model for an AVTD
spacetime necessarily has a non--oscillatory singularity, which excludes
generic spacetimes. 
The precise formulation of the BKL conjecture for generic
spacetimes is more subtle, 
see \cite{uggla:etal:G0,andersson:etal:gowdy} for 
discussion of this problem.

The numerical studies referred to above indicate that for a generic Gowdy
spacetime, as $\tau \nearrow \infty$, 
the velocity $\vhyp$ is eventually 
forced to satisfy $0 \leq \vhyp \leq 1$, except at isolated $x$--values, 
even if $\vhyp > 1$ in some subsets of $S^1$ initially, and further that $\vhyp$ has a
limiting value $\vhathyp(x)$ 
for each $x$ as one moves toward the singularity.

\begin{figure}[htb]
\centering
\includegraphics[width=3in]{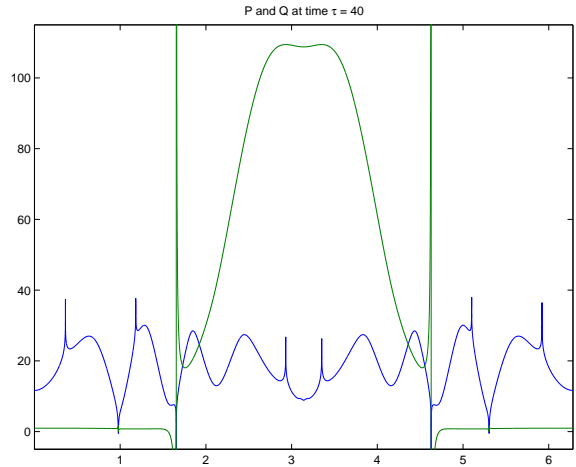}
\caption{Spikes in $P,Q$. The very sharp spikes (in $Q$), so--called
  ``false spikes'' are coordinate
effects.}. 
\label{fig:pqspikes}
\end{figure}

The numerical solutions exhibit ``spikes'' at those $x$--values, 
where $\vhyp \geq 1$ asymptotically, cf. figure \ref{fig:pqspikes}. The very
sharp spikes
seen (in $Q$) are coordinate effects, corresponding to a part of the solution
loop approaching the point on the boundary of $\bfH^2$ sent to infinity by
the transformation leading to the model with metric (\ref{eq:hypmet}).

\mnote{comment on artificial symmetry in figure \ref{fig:pqspikes}}
Gowdy spacetimes such that $\partial_x Q = 0$ 
are called \index{polarized}{polarized}. Equation (\ref{eq:gowdy-wavemap}) then becomes linear. 
It was proved by Isenberg and Moncrief \cite{isenberg:moncrief:asymptGowdy}
that polarized Gowdy spacetimes are AVTD. 
Numerical studies by Berger et. al., see 
\cite{berger:garfinkle:phenomenology} and references therein, see also 
\cite{hern:stewart:gowdy,berger:garfinkle:moncrief:revisit}, 
support the idea that general Gowdy spacetimes are AVTD.
It should be noted that there are polarized Gowdy spacetimes
with $v > 1$ up to the singularity, but the above mentioned 
work indicates that this behavior is non--generic. 
See also \cite{chrusciel:ANU} for work on the case $v = 0$. 

The equation (\ref{eq:gowdy-wavemap}) may, 
essentially under the restriction $0 < v < 1$, be written as a
\index{Fuchsian}{Fuchsian} system
$$
(t \partial t + E){\CMcal U} = {\CMcal F}[\CMcal U] ,
$$
where $t = e^{-\tau}$, 
and hence using a singular version of the Cauchy--Kowalew\-skaya theorem, cf. 
Kichenassamy and Rendall \cite{kichenassamy:rendall} and references therein,
see also \cite{baouendi:goulaouic}, AVTD solutions may be constructed given 
real analytic ``data on the singularity''. The solutions constructed are in
terms of the time coordinate $\tau$, of
the form 
\begin{subequations}\label{eq:Gowdy-AVTD}
\begin{align}
P(\tau,x) &= \vsign(x) \tau + \phi(x) + e^{-\eps\tau} u(x,\tau) \\
Q(\tau,x) &= q(x) + e^{-2\vsign(x) \tau} [ \psi(x) + w(\tau,x)]
\end{align}
\end{subequations}
where $\eps > 0$, $0 < v < 1$, and $w,u  \to 0$ as $\tau \to \infty$. 
The Fuchsian method
was generalized for the Gowdy case by Rendall \cite{rendall:fuchs-smooth} 
to the $C^\infty$ case.

Rendall and Weaver \cite{rendall:weaver:spikes}
have constructed families of solutions with spikes, starting from a solution 
with a false spike at $x_0$ with 
$ k = \vhathyp(x_0) < 1$, by applying an 
explicit transformation, a new solution with $\vhathyp(x_0) = 1 + k$. 
In the new
solution $\vhathyp$ has a discontinuity at 
$x_0$ with $\lim_{x \to x_0} \vhathyp(x)
= 1-k$. By iterating this procedure spikes with arbitrarily high velocity can 
be constructed. The group formed by composing the transformations used by
Rendall and Weaver has been termed the Geroch group \cite{chae:chrusciel}. 
Spikes with asymptotic velocity $\vhathyp > 2$ 
correspond to higher order
zeros of $\partial_x Q$, and are therefore non--generic. Garfinkle and Weaver
\cite{garfinkle:weaver:high}
have studied the dynamics of generic spikes with initally high
velocity. Their analysis, based on a combination
of heuristic arguments, related to the ``method of consistent potentials'',
see \cite{berger:garfinkle:phenomenology} for discussion, and numerical work, 
shows that the velocity
of these spikes is driven down into the interval $1 < v < 2$ by a sequence of
bounces, the qualitative features of which can be explained in terms of the
relative importance of the terms in the evolution equation.

Recently conditions on initial data (essentially a small
energy condition together with restriction on the velocity) 
have been given by Ringstrom
\cite{ringstrom:gowdy-vac, ringstrom:gowdy-exp}, and Chae and Chru\'sciel
\cite{chae:chrusciel} under which the solution is of the form 
(\ref{eq:Gowdy-AVTD}). This work shows that 
AVTD
behavior holds on open and dense subsets of $S^1$, and makes earlier work 
\cite{grubisic:moncrief:asymp} on formal expansions for the Gowdy field 
equations rigorous. The definition of AVTD solution used by Chae and
Chrusciel \cite[Eq. (3.8-3.9)]{chae:chrusciel} is more general than
(\ref{eq:Gowdy-AVTD}) 
used by Ringstrom, allowing more general velocity $\vsign$ but with 
less precise control on the lower order terms.

In view of the above, it is reasonable to make 
the following conjecture. 
\begin{conj} \label{conj:gowdyAVTD}
Generic vacuum, spatially compact $\U(1)\times \U(1)$--symmetric
spacetimes with vanishing twist 
are AVTD at the 
singularity, in the complement of an at most countable closed 
subset $E$ of $S^1$. 
The asymptotic velocity $\vhathyp(x)$ exists for all $x \in S^1$ and satisfies
$0 < \vhat(x)  < 1$ for $x \in S^1 \setminus E$. $\vhathyp$ is continuous on
$S^1 \setminus E$. 
\end{conj}
\begin{remark}
\begin{enumerate}
\item This is implicit in Grubi{\v{s}}i{\'c} and Moncrief \cite{grubisic:moncrief:asymp}.
\item Chae and Chru\'sciel \cite{chae:chrusciel} 
have constructed solutions with an asymptotic velocity which is discontinuous
on any closed set $F \subset S^1$, with nonempty interior. In particular $F$
may have nonzero measure. If the above
Conjecture is correct, this behavior is nongeneric. 
\end{enumerate}
\end{remark}
The question of cosmic censorship for the Gowdy spacetimes may be studied by
analyzing the behavior of the Kretschmann scalar $\kappa$ as $\tau \to
\infty$. 
For the class of polarized Gowdy spacetimes, 
this was done by Chru\'sciel et. al. \cite{chrusciel:SCC:gowdy}. 
\mnote{say what was done in \cite{chrusciel:SCC:gowdy}}
It was proved by Kichenassamy and Rendall  \cite{kichenassamy:rendall} 
that for generic AVTD spacetimes constructed using the Fuchsian algorithm, the 
Kretschmann scalar $\kappa$ blows up at the singularity and  
hence generically, these spacetimes do not admit extensions, 
see also the discussion in \cite[\S 3--4]{grubisic:moncrief:asymp}. The same
behavior was shown to hold also for the class of Gowdy spacetimes with
asympotic velocity $0 < \vhat < 1$ and bounds on the energy density, see
\cite{ringstrom:gowdy-exp,chae:chrusciel}. Further, the Gowdy spikes
constructed by Rendall and Weaver also have $\kappa$ blowing up, though at
a different rate than nearby points. 
Therefore a
reasonable approach to the SCC in the class of Gowdy spacetimes, is via
conjecture \ref{conj:gowdyAVTD}.

It is relevant to mention 
here that the AVTD behavior for Gowdy symmetric spacetimes
may be broken by the introduction of suitable matter, cf. 
\cite{isenberg:etal:inhomog}, where numerical evidence for an oscillatory
approach to the singularity is presented, for a locally $\U(1)\times \U(1)$
symmetric spacetime with magnetic field. Similarly, it is expected that
scalar field or stiff fluid matter changes the oscillatory behavior of
general $\U(1)\times \U(1)$ symmetric models to AVTD. However, even in the
presence of a scalar field, one expects to see spikes forming.

\mnote{discuss the consequences for curvature blowup of the work of
  Ringstrom and Chru\'sciel}

\mnote{mention special role of $v = 1$} 

A systematic way of deriving the system of ODE's governing the asymptotic
behavior of the Gowdy spacetimes is to write 
(\ref{eq:gowdy-wavemap}) in first order form, using the scale invariant
operators $\partial_\tau, e^{-\tau} \partial_x$, and then cancelling the terms
involving $e^{-\tau} \partial_x$. Following this procedure and working in terms
of $H$-normalized orthonormal frame variables adapted to the $\U(1)\times
\U(1)$ orbits, constructed in a completely
analogous way to those discussed in section \ref{sec:bianchi} 
gives, after cancelling the terms corresponding to $e^{-\tau}\partial_x$, the
system
\begin{subequations}\label{eq:hom-evol}
\begin{align}
%
\DD_0 \Sigp
& =  (q-2)\,(1+\Sigp) \\
\DD_0 \Sigm
& =  (q-2)\,\Sigm - 2\sqrt{3}\,\Nm^{2} + 2\sqrt{3}\,\Sigc^{2} \\
\DD_0 \Nc
& =  (q+2\Sigp)\,\Nc \\
\DD_0 \Sigc
& =  (q-2-2\sqrt{3}\Sigm)\,\Sigc - 2\sqrt{3}\,\Nc\,\Nm \\
\DD_0 \Nm
& =  (q+2\Sigp+2\sqrt{3}\Sigm)\,\Nm + 2\sqrt{3}\,\Sigc\,\Nc \ .
\end{align}
\end{subequations}
where 
\mnote{check def of $\DD_0$} 
$\DD_0 = (1+\Sigp) \partial_\tau$, 
$q = 2(\Sigp^{2}+\Sigm^{2}+\Sigc^{2})$, 
subject to the constraint equation
\begin{align}
1 & =  \Sigp^{2}+\Sigm^{2}+\Sigc^{2}+\Nc^{2}+\Nm^{2} 
\label{eq:hom-con}
%
\end{align}
\mnote{fix the figures to look similar}
\begin{figure}[!tbp]
\begin{minipage}[t]{0.45\linewidth}
    \centering
    \includegraphics[width=2in]{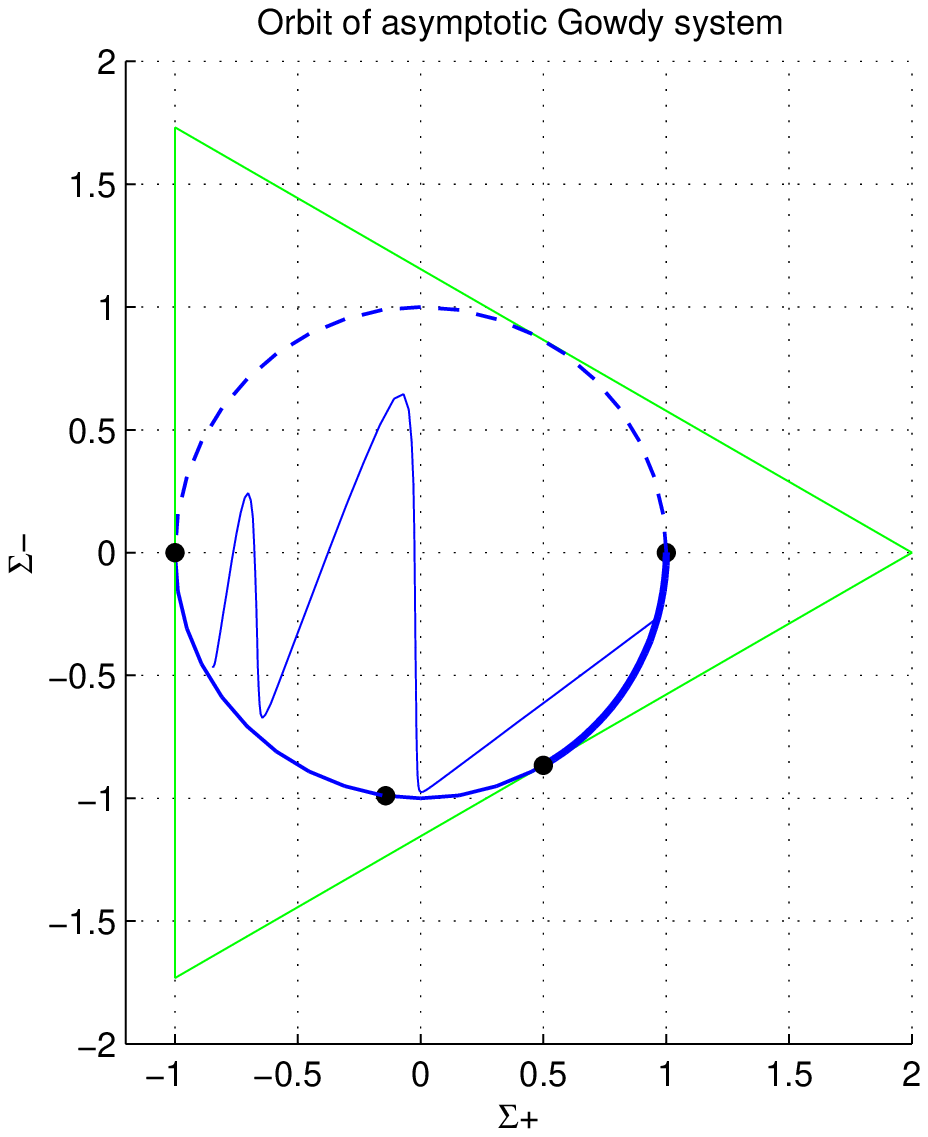}
    \caption{Projection in Kasner plane of an orbit of the asymptotic Gowdy 
    system 
    (\ref{eq:hom-evol}--\ref{eq:hom-con}).}
    \label{fig:gowdyasympt-proj}
\end{minipage}%
\begin{minipage}[t]{0.45\linewidth}
    \centering
\includegraphics[width=2in]{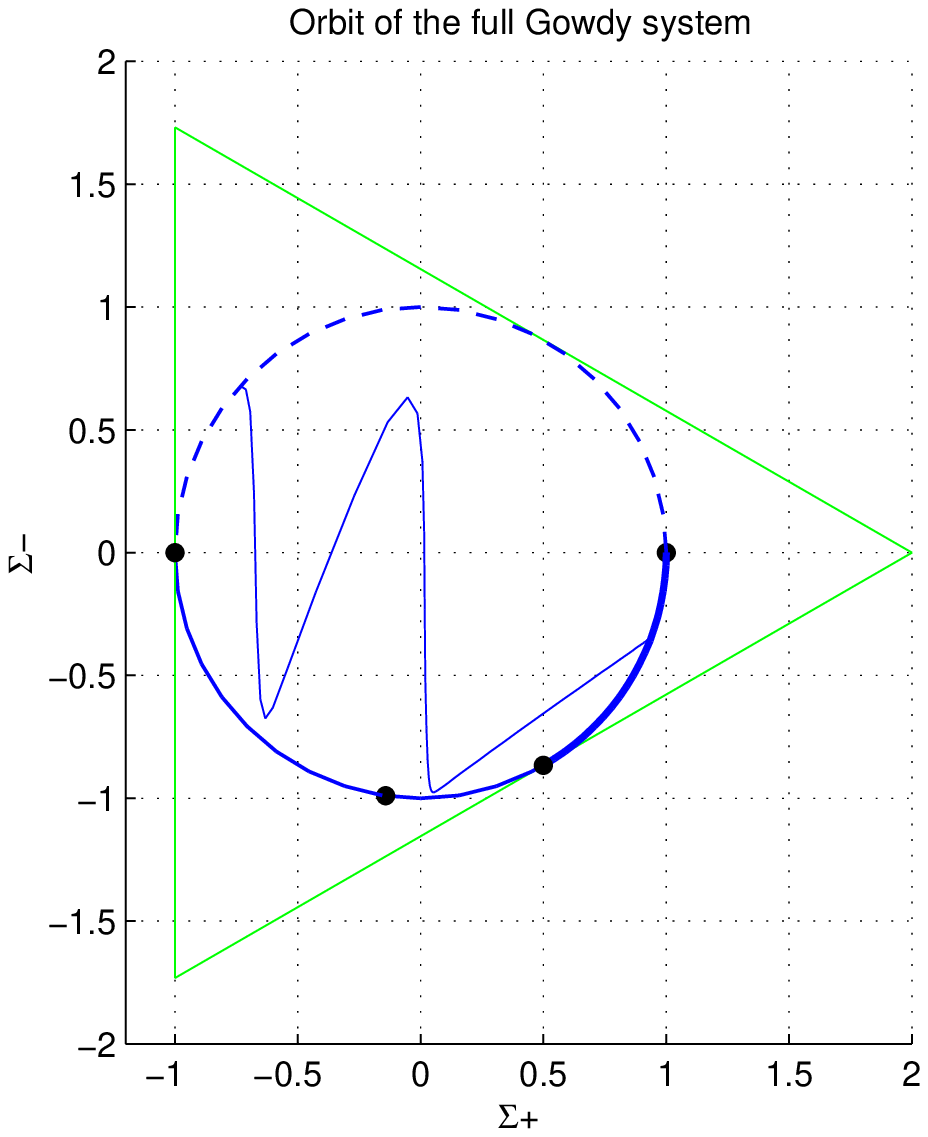}
    \caption{Projection in Kasner plane of the orbit of a point of the full
    Gowdy system. }
    \label{fig:spikepoint-fulldyn-proj}
\end{minipage}%
\end{figure}
See \cite{andersson:etal:gowdy} for discussion of the system
(\ref{eq:hom-evol}--\ref{eq:hom-con}).  This system
contains the spatially homogenous Bianchi class A models I--VII$_0$, but
represented in terms of a non--Fermi--propagated frame.

The asymptotic behavior of the system (\ref{eq:hom-evol}--\ref{eq:hom-con})
in the direction towards the singularity 
is easily analyzed, and one shows that all solutions of the system have limit
points on the Kasner half--circle $\Sigma_+^2 + \Sigma_-^2 = 1$. Similar to
the standard Bianchi case,
the Kasner circle consists of fixed points. For the Bianchi types
II--VII$_0$, only part of the Kasner circle consists of stable fixed
points. In the present case, only the arc with $\Sigma_- \leq 0$, $\Sigma_+ > \half$ is stable, and
a generic orbit of (\ref{eq:hom-evol}) ends on the stable arc. The
stable arc corresponds to orbits with asymptotic velocity $0 < \vhat < 1$. 
See figure \ref{fig:gowdyasympt-proj} for an example of an orbit of the system
(\ref{eq:hom-evol}--\ref{eq:hom-con}). 

According to the BKL proposal, a generic solution $u(t,x)$ of the full Gowdy 
system will have the property that it will asymptotically approach a solution
of the Gowdy AVTD system (\ref{eq:hom-evol}--\ref{eq:hom-con}). That this is
indeed the case is indicated by figure \ref{fig:spikepoint-fulldyn-proj}, 
which shows the
projection in the Kasner plane of the 
orbit of one point for the full Gowdy system. We see in the figure the orbit
of a point near a spike point. The points marked on the Kasner circle are
(from the left) the points with velocity $v = \infty, 2, 1, 0$. The stable
arc on the Kasner circle is plotted with a heavy line. 
It should be noted that the last part of the orbit which ends on the stable
Kasner arc is approximately a
Bianchi II orbit, with initial velocity in the interval $(1,2)$. This is the
generic behavior, also close to spike points, 
see \cite{andersson:etal:gowdy} for more details.

A class of AVTD
polarized Gowdy spacetimes with non--vanishing twist has been constructed by
Isenberg and Kichenassamy \cite{isenberg:kichenassamy:gowdy} using the
Fuchsian algorithm. On the other hand for general non--polarised twisted $\U(1)
\times \U(1)$ symmetric spacetimes, work of Berger et
al. \cite{berger:etal:T2oscill} 
supports the conjecture that these spacetimes show
oscillating behavior as one approaches the singularity.
This idea is also supported by the fact that the only Bianchi model with
$\U(1)\times \U(1)$ symmetry and non--vanishing twist constants is the
exceptional Bianchi VI$_{-1/9}$ which has been shown heuristically by 
Hewitt et al.
\cite{hewitt:etal:VI-1/9} to have oscillatory behavior at the singularity.

\mnote{note distinction between SH version and ``asymptotic dynamical
  system'', it is not obvious this gives the same system for general $G_2$!!,
  see commented out text for another version}
\mnote{add ref to Isenberg \& Weaver paper on $\U(1)\times \U(1)$ with twist}


\mnote{mention future asympot for Gowdy (Ringstrom) and other similar results} 

\mnote{mention Moncrief and Tanimoto, pert. of Bianchi etc}

\section{$\U(1)$}\label{sec:U(1)}
\mnote{check notation to conform to the ChB-Moncrief paper}
The $\U(1)$ symmetric vacuum 3+1 Einstein equations is an important case
which is of intermediate difficulty between the full 3+1 Einstein equations
and the highly symmetric Gowdy equations. 
In the presence of a hypersurface orthogonal space--like
Killing field, the Einstein equations reduce to $2+1$ gravity coupled to
wave map matter, the field equations and their reductions
have been derived in
\cite{moncrief:U(1),moncrief:U(1):annphys,moncrief:U(1):EMH}, see also \cite{moncrief:cameron}.
Choquet--Bruhat and Moncrief have proved that the Cauchy problem for the
$\U(1)$ problem is well--posed in $H^2$, see 
\cite{ChBM:U(1):global,ChB:U(1):global}.
\index{$\U(1)$ symmetric spacetime}{}

In the above quoted papers, the spacetime is assumed to be a $\U(1)$ bundle
over a spatially compact $2+1$ spacetime. The
case of local $\U(1)$ symmetry does not appear to have been considered in
connection with the Einstein equations, see however
\cite{orlik:raymond:localSO(2)} for information about 3--manifolds with local
$\U(1)$ action. 

It is also possible to study the
case when the reduced space is asymptotically flat. This case has been
considered in work by Ashtekar and others, 
\cite{ashtekar:varadarajan,ashtekar:bicak:schmidt:asymp}, see also 
\cite{ashtekar:bicak:schmidt:null}. 
In these papers an analogue of the 
ADM mass at spatial infinity is introduced. It is
proved that it is nonnegative and bounded from above. It is
interesting to study the consequences of the presence of this conserved 
quantity for the 2+1 dimensional 
Einstein--matter system given by the $\U(1)$ 
problem, as one expects that it gives a stronger bound on the fields
than in the 3+1 case. This appears to be a natural setting for a small
data version of the $\U(1)$ problem. 

In the following, we will consider the spatially compact case.
Let $(\aM,\ame)$ be a 3+1 dimensional spacetime, assume 
$\aM \cong B \times \Re$, with $\pi: B \to \Sigma$ a principal $\U(1)$
bundle, $\Sigma$ a compact surface. 
Further assume the group $\U(1)$ acts by isometries on $(\aM, \ame)$
with the action generated by the Killing field $J$, which we assume to be
space--like, $\la J, J \ra > 0$. 

Let the function $\lambda$ on $\Sigma \times \Re$ be defined by 
$\pi^* \lambda = \half \log (\la J , J \ra)$, and let 
$\theta = e^{-2\pi^*\lambda} J$. 
Then we can write the spacetime metric $\ame$ in the form 
$$
\ame = \pi^* (e^{-2\lambda} g) + e^{2\pi^*\lambda} \theta \tens \theta ,
$$
where $g$ is a Lorentzian metric on $\Re \times \Sigma$. 

Introduce a 
frame\footnote{In \cite{ChB:moncrief:U(1)}, a frame $dx^a, \theta^3$ is
used, where 
$\theta^3_{\alpha} = e^{-2\lambda} J_{\alpha} = e^{-\lambda} e^3_\alpha $ .
Furthermore, their $F_{\alpha\beta} = d\theta_{\alpha\beta} = e^{-2\lambda}
\Theta_{\alpha\beta}$.}
$e_{\alpha}$, $\alpha = 0,1,2,3$ with $e_3 = e^{-\lambda} J$, and let the
indices $a,b,c,\dots = 0,1,2$. We may without loss of generality assume
that 
$
[J, e_a ] = 0 .
$
We have 
$$
dJ_{\alpha\beta} = \Theta_{\alpha\beta} 
+ 2 ( e_{\alpha}( \lambda) J_{\beta} -  e_{\beta}(\lambda) J_{\alpha} ) ,
$$
where $\Theta_{\alpha\beta} J^\beta = 0$. It follows that 
$\Theta_{\alpha\beta} = \pi^* (e^{2\lambda} F_{\alpha\beta})$ 
where $F_{\alpha\beta}$ is a 2--form on 
$\Re \times \Sigma$. If $F_{\alpha\beta} = 0$, then 
$J$ is hypersurface orthogonal. 

To avoid cluttering up the notation we will in the following 
make no distinction between fields on the orbit space $\Re \times B$ and
their pullbacks by $\pi$.

The components of the 3+1 Ricci tensor are 
\begin{align}
\aR_{ab} &= R_{ab} - 2 \nabla_a \lambda \nabla_b \lambda 
+ \nabla_c \nabla^c \lambda g_{ab} - \half e^{4\lambda} F_{ac} F_a^{\ c} ,
\nonumber \\
\aR_{a3} &= \half e^{-\lambda} \nabla_c (e^{4\lambda} F_a^{\ c}) , \nonumber
\\
\aR_{33} &= e^{2\lambda} [ - g^{ab} \nabla_a \nabla_b \lambda 
+ \quart e^{4\lambda} F_{ab} F^{ab} ] \label{eq:aR33}
\end{align}

Let the one--form $E$ on $\Re \times \Sigma$ be given by 
$E = - \star_g (e^{4\lambda} F)$. One of the Einstein equations 
(\ref{eq:aR33}) implies that $dE = 0$. 

%

Now the Einstein vacuum equations $\aR_{\alpha\beta} = 0$ imply the
system 
\begin{subequations}\label{eq:U(1)einst}
\begin{align}
R_{ab} &= \half ( 4 \nabla_a \lambda \nabla_b \lambda + e^{-4 \lambda} E_a
E_b),  \label{eq:geq}\\
\nabla^a \nabla_a \lambda 
+ \half e^{-4\lambda} E_a E_b
g^{ab} &= 0 , \label{eq:lambdaeq}\\
\nabla^a \nabla_a \omega - 4 \nabla^a \lambda E_a &=0 .
\label{eq:omegaeq}
\end{align}
\end{subequations}

When $E_a = \nabla_a \omega$, we recognise (\ref{eq:U(1)einst}) as the 2+1
dimensional Einstein equations with
\index{wave map}{wave map} matter, for the wave map with components $(\lambda, \omega)$ with
target hyperbolic space $\bfH^2$ with the constant curvature metric 
$$
2 d\lambda^2 + \half e^{-4\lambda} d\omega^2 .
$$
See subsection \ref{sec:PDE} for some discussion of wave map equations. 

We now specialise even further to the polarized case $E_a = 0$. 
This corresponds to assuming
that the bundle $\pi: B \to \Sigma$ is trivial, and that the vector field 
$J$ is hypersurface orthogonal. Then $\omega$ is constant and 
the equations (\ref{eq:U(1)einst})
become 
\begin{subequations}\label{eq:U(1)polar}
\begin{align}
R_{ab} &= \nabla_a \lambda \nabla_b \lambda \\
\nabla^a \nabla_a \lambda &= 0
\end{align}
\end{subequations}
which is precisely the 2+1 Einstein equations coupled to a massless scalar
field. 

In the 2--dimensional case, the operator $k \to \div k$ is elliptic on
symmetric 2--tensors with vanishing trace. 
Further, by the uniformization theorem, 
a compact 2--dimensional Riemannian manifold $(\Sigma, h)$ is conformal to 
$(\Sigma, [h])$ where $[h]$ is a representative of the conformal class of
$g$, i.e. a constant curvature metric. Therefore
working in spatial harmonic gauge with respect to 
$(\Sigma, [h])$ (i.e. conformal spatial harmonic gauge) 
and CMC time gauge, the constraint
equations form an elliptic system for $(h_{ij}, k_{ij})$. We get a
representation of $(h_{ij}, k_{ij})$ in terms of $([h], \phi, k^{\TT}, Y)$
where $[h]$ is the conformal class of $h$, corresponding to a point in
Teichm\"uller space, $\phi$ a conformal factor determined by the Hamiltonian
constraint equation (a nonlinear elliptic system for $\phi$), 
$k^{\TT}$ a trace--free, divergence free 2--tensor
on $(\Sigma, [h])$, corresponding to a quadratic differential, 
and finally,  $Y$ is a vector field determined from the 
momentum constraint equation. 
Note that 
$([h],k^{\TT})$ represents a point in $T^*\Teich(\Sigma)$, 
the cotangent bundle of the Teichm\"uller space of $\Sigma$. 

Due to the ellipticity of the constraint equations in the 2+1 dimensional 
case, in the gauge as described above, it is possible to eliminate the 
Einstein equation (\ref{eq:geq}) from the system (\ref{eq:U(1)einst}) and
instead solve the elliptic--hyperbolic system consisting of the hyperbolic 
system (\ref{eq:lambdaeq}-\ref{eq:omegaeq}), coupled to a nonlinear system
which determines components of the spacetime metric $\ame$ in terms of the
data via the constraint and
gauge fixing equations, as well as an ODE system determining the evolution of
the Teichmuller degrees of freedom of the metric on $\Sigma$.  In the special 
case $\Sigma = S^2$, Teichmuller space is a point, and further $H^1(S^2)=0$,
which means that $E = d\omega$.  Therefore ignoring problems with gauge fixing,
the reduced $\U(1)$ system would consist of the wave equation 
coupled to the system which
determines the components of $\ame$. 

A special case of the \index{polarized}{polarized}
$\U(1)$ equations is given by setting
$\lambda \equiv$ constant. Then the field equations (\ref{eq:U(1)polar})
are just 
$$
R_{ab} = 0 ,
$$
the 2+1 dimensional vacuum equations. In this case, the spacetime is a
3-dimensional Lorentzian space--form. The dynamics of 2+1 dimensional 
vacuum gravity has been studied by Andersson, Moncrief and Tromba
\cite{andersson:etal:2+1grav} who proved global existence in CMC time for
$2+1$ dimensional vacuum spacetimes, with cosmological constant,  
containing at least one CMC hypersurface. This
proves Conjecture \ref{conj:CMC} for the class of 2+1 dimensional 
vacuum spacetimes. 

Andersson and Rendall \cite{andersson:rendall:quiescent} showed that the 3+1
dimensional Einstein scalar field system can be formulated as a
\index{Fuchsian}{Fuchsian}
system. Using similar techniques, 
Damour et al. \cite[\S 4]{damour:etal:kasnerlike} 
have shown 
that the polarized $\U(1)$
equations may be formulated as a Fuchsian system, 
and therefore AVTD solutions
may be constructed using a singular version of the Cauchy-Kowalewskaya
theorem, as was done by Kichenassamy and Rendall \cite{kichenassamy:rendall}
for the Gowdy case. See also 
Isenberg and Moncrief \cite{isenberg:moncrief:U(1)fuchs} which deals with the
``half--polarized'' $\U(1)$ system in addition to the polarized case.  
This supports the following conjecture. 
\begin{conj} \label{conj:U(1)AVTD}
Generic polarized $\U(1)$ spacetimes are AVTD at 
the singularity. 
\end{conj}
\begin{remark} Conjecture \ref{conj:U(1)AVTD}
was essentially stated by Grubi{\v{s}}i{\'c} and Moncrief
\cite{grubisic:moncrief:mixmaster}.
This is supported by numerical work of Berger and Moncrief
\cite{berger:moncrief:U1polarized}. Similarly to the case of Gowdy 
spacetimes, see
Conjecture \ref{conj:gowdyAVTD}, one
expects that ``spiky features'' will form at the singularity, and hence the
AVTD property of the $\U(1)$ symmetric spacetimes should be understood to be
generic also in the spatial sense.
\end{remark}

It seems reasonable to expect that polarized 
$\U(1)$ spacetimes which are AVTD at the
singularity have a strong curvature singularity generically, and therefore
that proving the AVTD conjecture for polarized $\U(1)$ would be a big step
towards proving SCC for this class. 

In contrast to polarized $\U(1)$, the generic $\U(1)$ spacetimes have
sufficiently many degrees of freedom that one expects them to satisfy the
BKL picture of an oscillatory approach to the singularity, as is also
expected in the fully 3+1 dimensional case, cf. the remarks in section
\ref{sec:concluding}. This is supported by the numerical evidence so far, see 
\cite{berger:moncrief:U1oscillatory}. 

In the expanding direction, on the other hand, a small data semi--global 
existence result holds \cite{ChBM:U(1):global,ChB:U(1):global}, 
similar to the one for the full 3+1 dimensional case discussed below in section
\ref{sec:3+1}. In this case, the notion of ``small data'' is taken to be
data close to data for a background spacetime with space--like slice 
$M = \Sigma \times S^1$ where $\Sigma$ is a Riemann surface of genus $>2$ with 
metric $\sigma$ of scalar curvature $-1$, 
and $M$ has the product metric. In this case, the 
background spacetimes are
products of flat spacetimes as in Example \ref{ex:misner} with $n=2$, with
the circle (such spacetimes are
of type Bianchi III). The proof uses energy estimates for a second order
energy, which controls the $H^2 \times H^1$ norm of the data, for small
data. The energy expression used is a combination of several terms. The
first order part is $\frac{\tau^2}{4} \Vol_g(\Sigma) + 2\pi \chi(\Sigma)$,
which via the constraint equations may be related to the wave map energy. The
second order term is defined in terms of the $L^2$ norm of the Laplacian of
the wave map field, and the gradient of its time derivative. In order to use
this method, one of the technical problems one has to deal with is
controlling the spectral gap of the Laplacian. This is done by keeping
control of the Teichmuller parameters during the evolution. For technical
reasons, the assumption is made 
that the spectral gap $\Lambda_0$ of the Laplacian
for the inital metric satisfies $\Lambda_0 > 1/8$. The proof shows that the
Teichmuller parameter $\sigma(t)$ converges to a point in the interior of
Teichmuller space and that the spacetime geometry converges in the expanding
direction to one of the model spacetimes described above. 

\mnote{add ref to Damour et al gr-qc/0202069, one result \S 4 concerns 4d
Einstein eqs with one KV} 

\mnote{mention Piotr's work on Robinson-Trautman and Alan's smalll data R-T
work} 

\mnote{mention flat spacetimes, 2+1 asymptotics etc}  

\section{3+1}\label{sec:3+1}
In the 3+1 dimensional case with no symmetries,
the only known facts on the global properties
of spacetimes are Lorentzian geometry results such as the Hawking--Penrose
singularity theorems 
\cite{penrose:difftop}
and the Lorentzian splitting theorem of Galloway \cite{galloway:complete}, 
see \cite{beem:ehrlich:global} for a survey of Lorentzian geometry. 

Here we are interested in results relevant to the SCC, Conjecture 
\ref{conj:SCC} and the CMC conjecture, Conjecture \ref{conj:CMC},
i.e. results about the global behavior of solutions to the Cauchy problem
for the evolution Einstein equations, in some suitable gauge. 
The Cauchy problem for the Einstein evolution equations has been discussed in 
section \ref{sec:evoleqs}. 

With this limitation there are essentially 
3 types of results known and all of 
these are small data results. The results are those of Friedrich on the 
``hyperboloidal Cauchy problem'', of Christodoulou and Klainerman on the
nonlinear stability of Minkowski space (generalized by 
Klainerman and Nicol\`o \cite{klainerman:nicolo:book} to exterior domains) 
and the work of Andersson and Moncrief \cite{andersson:moncrief:global},
on global existence to the future for data close to the data for certain
spatially compact flat $\kappa =-1$ (local) FRW spacetimes, again a nonlinear
stability result. We will briefly discuss the main features of these results. 

The causal structure of a Lorentz space is a conformal invariant. This leads 
to the notion that the asymptotic behavior of spacetimes can be studied
using conformal compactifications or blowup. 
The notion of isolated system in general relativity has been formalized by
Penrose in terms of regularity properties of the boundary of a conformally
related spacetime $(\tM, \tme)$, with null boundary $\Scri$, such that 
$\tM$ is a completion of $\aM$, and $\ame =
\Phi^* (\Omega^{-2} \tme)$, where $\Omega \in C^{\infty} (\aM)$ is a conformal 
factor, $\Phi: \aM \to \tM$ is a diffeomorphism of $\aM$ to the
interior of $\tM$. 
Given assumptions on the geometry of $(\tM, \tme)$ at $\Scri$, Penrose
proved using the Bianchi identities that the components of the Weyl tensor
of $(\aM, \ame)$ decay at physically reasonable rates.

Friedrich derived a first order symmetric
hyperbolic system from the Einstein equations, the ``regular conformal field 
equations''. This system includes among its unknowns, components of the Weyl 
tensor, the conformal factor $\Omega$ and quantities derived from the
conformally rescaled metric $\tme$. 
This system has the property that under the Penrose regularity conditions
at $\Scri$, the solution can be extended across $\Scri$.
The fact that the regular conformal field equations gives a
well posed evolution equation in the conformally compactified picture
enabled Friedrich to prove small data global existence results by using the
stability theorem for quasi--linear hyperbolic equations. 

In \cite[Theorem 3.5]{friedrich:complete} Friedrich proved 
global 
existence to the future for data $(M,g,k)$ close to the standard data on a
hyperboloid in Minkowski space, satisfying asymptotic regularity conditions
compatible with a Penrose type compactification (the hyperboloidal initial
value problem). This was later generalized to 
Maxwell and Yang--Mills matter in 
\cite{friedrich:EYM}. 
Initial data for the hyperboloidal initial value problem were first
constructed by Andersson, Chru\'sciel and Friedrich
\cite{andersson:etal:yamabe}, see also \cite{andersson:chrusciel:PRL},
\cite{andersson:chrusciel:cauchy}. 

\begin{figure}[htb]
\centering
\input{friedrich.eepic}
\caption{The semi--global existence theorem of H. Friedrich}
\label{fig:friedrich}
\end{figure}
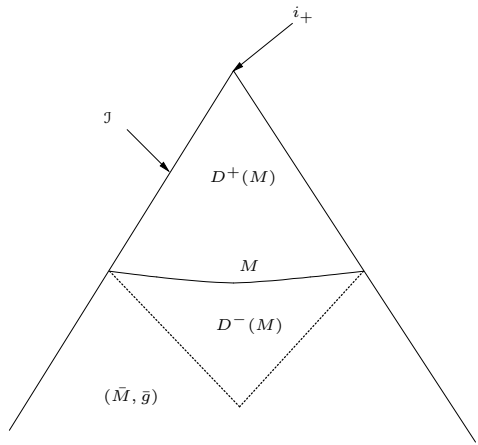

The result of Friedrich is a semi--global existence result, in the sense that
the maximal vacuum Cauchy development $D(M)$ of the data 
$(M,g,k)$ is proved to be geodesically
complete and therefore inextendible to the future, but not to the past. 
In fact typically $D(M)$ will be extendible to the past and $(M,g,k)$ may be 
thought of as a partial Cauchy surface in a larger maximal globally hyperbolic
spacetime $(\aM, \ame)$, cf. figure \ref{fig:friedrich}. In view of the
inextendibility to the future of $D(M)$, the result of Friedrich may be viewed 
as supporting the cosmic censorship conjecture. In the case of the Einstein
equations with positive cosmological constant, the method of Friedrich yields 
a global existence result for data close to the standard data on $M = S^3$ in 
deSitter space, cf. \cite[Theorem 3.3]{friedrich:complete}.

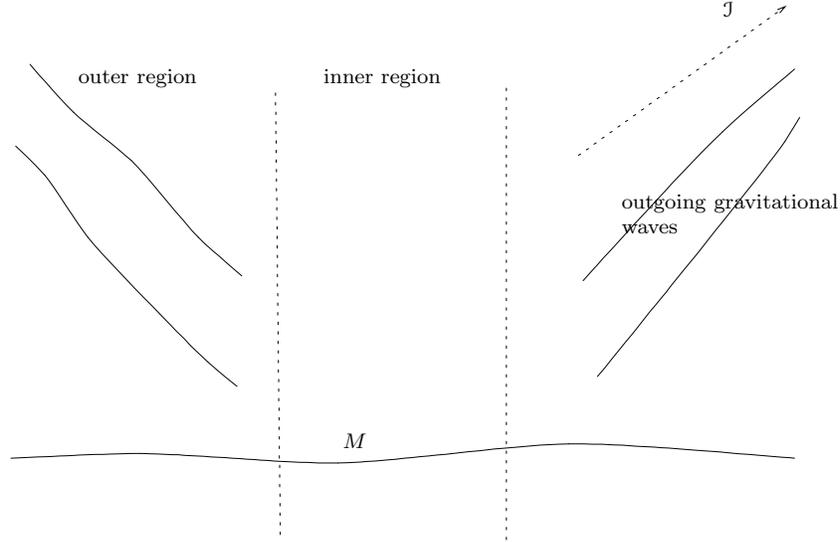
\begin{figure}[htb]
\centering
\input{christo.eepic}
\caption{The global existence theorem of D. Christodoulou and S. Klainerman}
\label{fig:christoklain}
\end{figure}

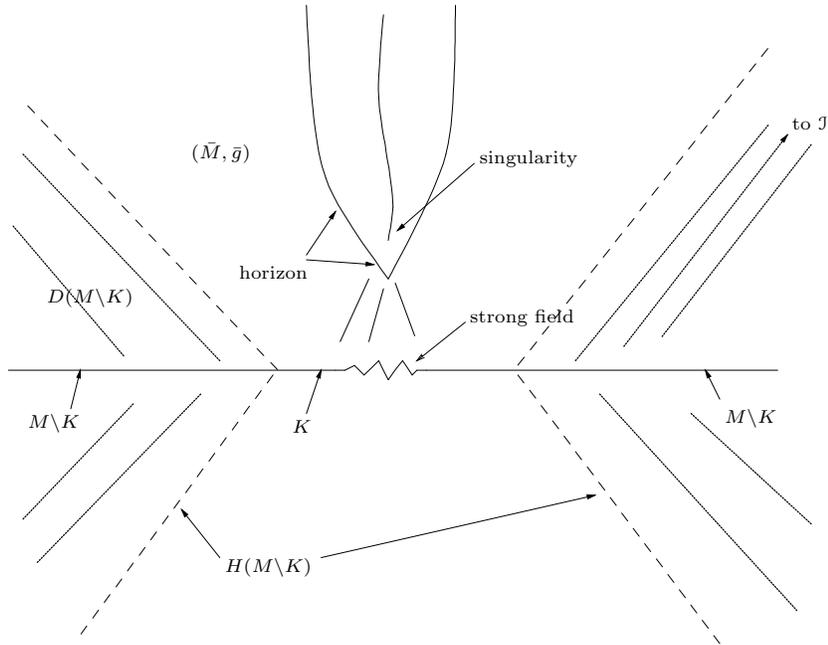
\begin{figure}[htb]
\centering
\input{ckn.eepic}
\caption{The exterior global existence 
theorem of S. Klainerman and
F. Nicol\'o. The figure illustrates a situation which is 
covered by this theorem, with 
a singularity forming due to a strong gravitational field in the interior 
region $I(K)$ while the exterior
region $D(M\setminus K)$ has complete outgoing null geodesics reaching
$\Scri$.}
\label{fig:KN}
\end{figure}

The first true small data global existence result for the vacuum Einstein
equations, was proved by D. Christodoulou and S. Klainerman
\cite{christo:klain:stability}. 
They proved that for data $(M,g,k)$ sufficiently close to standard data on a
hyperplane in Minkowski space, with appropriate decay at spatial infinity,
the MVCD $(\aM, \ame)$ is geodesically complete and therefore inextendible, 
cf. figure \ref{fig:christoklain}. 
The Christodoulou--Klainerman 
global existence theorem therefore supports the cosmic censorship conjecture.
Klainerman and Nicol\`o
\cite{klainerman:nicolo:review,klainerman:nicolo:book}
have proved a global existence result 
for data $(M,g,k)$ which are close
to standard flat Minkowski space data on an 
exterior region $M \setminus K$, where $K \subset M$ is compact so that 
$M \setminus K \cong \Re^3 \setminus \text{Ball}$, and with weaker asymptotic
conditions compared to the Christodoulou--Klainerman theorem.
The result of 
Klainerman and Nicol\'o states that for a vacuum data set
$(M,g,k)$ which is sufficiently close to the standard flat Minkowski data on
$M \setminus K$, the outgoing null geodesics in the causal exterior region 
$D(M\setminus K)$ are complete and $D(M \setminus K)$ is covered by a double
null foliation, with precise control over the asymptotics. It should be noted 
that the KN theorem therefore covers a more 
general class of
spacetimes than the Christodoulou--Klainerman theorem and is not strictly a
small data result. 
In particular, the maximal Cauchy vacuum development of $(M,g,k)$ may be
singular for data satisfying the assumptions of the KN theorem, cf. figure
\ref{fig:KN}. If the
smallness assumption is extended also to the interior region, the
Christodoulou--Klainerman theorem is recovered. 
The Einstein equations are quadratic in first derivatives,
and therefore in 3+1 dimensions, one needs something like a 
null condition in order to get
sufficient decay for a global existence argument. The Einstein equations are 
not known to satisfy a null condition.  However, by a detailed
construction of approximate Killing fields and approximate conformal Killing 
fields, Christodoulou and Klainerman are able to control the behavior of 
components of Bel--Robinson tensors 
constructed from the Weyl tensor and its derivatives, and 
close a bootstrap argument which gives global existence for sufficiently 
small data. As part of this argument, it is necessary to get detailed control
over the asymptotic behavior of light cones. 
Christodoulou and Klainerman also study the asymptotic behavior of
components of the Weyl tensor and are able to prove that some, but not all
of these have the decay implied by the Penrose conditions on $\Scri$. 

The question about existence of vacuum asymptotically Minkowskian
spacetimes with regular conformal completion, in the sense of Penrose, has
been open for a long time. 
Recently, Delay and Chru\'sciel \cite{delay:chr:simple}
were able to use a gluing construction of Corvino and Schoen
\cite{corvino:gluing} to construct vacuum spacetimes which are exactly
Schwarzschild near spatial infinity, and which are close enough to Minkowski
space, so that the global existence result of Friedrich applies to show
existence of a global $\Scri$. 
\mnote{is this future $\Scri$ or both past and future?} 
Further, Klainerman and Nicol\'o \cite{klainerman:nicolo:peeling}
have been able to show that a class of
asympotically Euclidean initial data with suitable decay at infinity has a
Cauchy development where the null components of the Weyl tensor have decay
corresponding to the peeling conditions of Penrose. 

The work in \cite{andersson:etal:yamabe} and
\cite{andersson:chrusciel:PRL} shows that generic hyperboloidal data do not
satisfy the required regularity at the conformal boundary, which therefore
may be viewed as an indication that Penrose regularity at $\Scri$ is
non--generic. It is still an open question what general conditions on
initial data on an asymptotically flat Cauchy surface give a Cauchy
development with regular conformal completion. H. Friedrich has been
developing a programme which approaches this problem by analyzing the
conformal structure at spatial infinite in detail, see 
\cite{dain:friedrich:AFreg} for references on this, and see also the paper 
\cite{valientekroon:obstruct} by Valiente-Kroon, which points out some new
obstructions to regularity.   

In contrast to the results of Friedrich and
Christodoulou--Klainerman--Nicol\'o, the 
semi--global existence result of Andersson 
and Moncrief \cite{andersson:moncrief:global} deals with spatially compact
vacuum spacetimes. 

Let $(M,\gamma)$ be a compact hyperbolic 3-manifold with metric $\gamma$ of
sectional curvature $-1$. Then the spacetime $\aM = M \times \Re$ with metric 
$\agamma = -dt^2 + t^2 \gamma$ is a  $\kappa = -1$ (local) 
FRW spacetime. A flat spacetime metric on $\aM$ defines a geometric structure
with group $SO(3,1)\ltimes \Re^4$. The moduli space of flat spacetimes with
topology $M \times \Re$ has the same dimension as the moduli space of flat
conformal structures on $M$, see \cite{andersson:flatCMC}. We say that $M$ is
{\bf rigid} if the dimension of the moduli space of flat conformal structures
on $M$ is zero. This is equivalent to the condition that there are no nonzero
traceless Codazzi tensors on $(M,\gamma)$. Kapovich \cite{kapovich:deform}
has constructed examples of rigid compact hyperbolic 3-manifolds. 

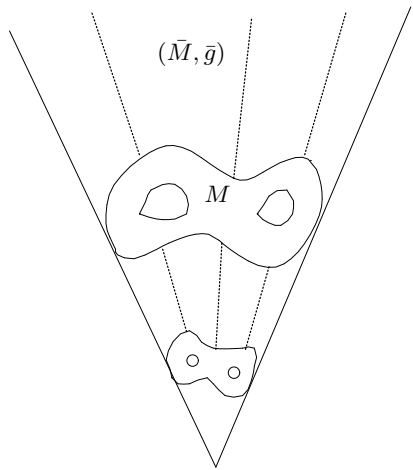
\begin{figure}[htb]
\centering
\input{am.eepic}
\caption{The semi--global existence theorem of L. Andersson and V. Moncrief.}
\label{fig:AM}
\end{figure}

The global existence theorem proved in \cite{andersson:moncrief:global} 
states that if $M$ is a rigid manifold of hyperbolic type, then 
for a vacuum data set $(M,g,k)$, 
sufficiently close to the 
standard data in a spatially compact $\kappa = -1$ (local) 
FRW spacetime, the 
MVCD $(\aM, \ame)$ is causally geodesically complete in the expanding
direction. It is a consequence of the singularity theorem of Hawking and
Penrose that $(\aM, \ame)$ is singular, i.e. geodesics in the collapsing
direction are incomplete, cf. figure \ref{fig:AM}. 

The proof uses local well posedness and a continuation principle
for the Einstein equations in constant mean curvature and spatial harmonic
coordinates (CMCSH) gauge, cf. subsection \ref{sec:CMC}, 
together with energy estimates for a first order 
Bel--Robinson type energy, defined in terms of the Weyl tensor and its
covariant derivative w.r.t. $T$. Let 
$W_{abcd}$ be the Weyl tensor and let $^*W_{abcd}$ be its dual. 
In a vacuum spacetime $W_{abcd} = R_{abcd}$. Define $E_{ij} = W_{iTjT}$ and
$B_{ij} = {}^*W_{iTjT}$. Then $E_{ij}$ and $B_{ij}$ are the electric and
magnetic parts of the Weyl tensor. These fields are traceless and symmetric
and satisfy a Maxwell--like
system. A computation shows that the zeroth order Bel--Robinson energy  
$$
\EBR = \int  |E|^2 + |B|^2
$$
(which measures second derivatives of $g$ in $L^2$),
satisfies the equation 
\begin{align*}
\partial_t  &= 
- 3 \int_{M_t} \Lapse [(E\times E)\cdot k  + (B\times B)\cdot k  
 - \third (|E|^2 + |B|^2)\tr k \\
&\quad - 2 \Lapse^{-1} \nabla^i \Lapse  (E \wedge B)_i ] d\mu_{M_t} 
\end{align*}
where for symmetric traceless two-tensors $A,B$,  
$$
(A \times B)_{ij} = A_i^{\ m}B_{mj} +  A_j^{\ m} B_{mi} 
- \frac{2}{3}  A^{mn} B_{mn} g_{ij} 
$$
and 
$$
(A \wedge B)_i = \Eps_i^{\ mn} A_m^{\ k}B_{kn} ,
$$
Combining $\EBR$ with the a term defined analogously in terms of $\anabla_T
W_{abcd}$ and extracting a scale invariant quantity, leads to an energy
$\Energy$, which bounds the Cauchy data in $H^3\times H^2$ for small data.
This energy is shown to satisfy an energy inequality of the form 
$$
\partial_t \Energy  \leq - 2\Energy - C \Energy^{3/2}
$$
Using this one may show that if $\Energy$ is small initially, then $\Energy$
will decrease, as long as local existence holds. A continuation principle for
the Einstein equations in $H^3\times H^2$ \cite{andersson:moncrief:local}
completes the proof of global existence for small data.

\section{Concluding remarks}\label{sec:concluding}
The results and numerical studies so far can be argued to fit with the so
called \index{BKL}{BKL} 
picture of cosmological singularities, which states that the 
generic singularity should be space--like, local and oscillatory, see
\cite{berger:etal:generic} for a review. Roughly, one 
expects that the ``points on the singularity'' are causally separated and
that locally at the singularity, the dynamics undergoes a chaotic sequence of
curvature driven ``bounces'' interspersed with relatively uneventful
``coasting'' epochs, 
with Kasner like dynamics (i.e. the metric is, locally in space, 
approximately of the form
given in (\ref{eq:kasner})). The locality at the singularity, 
is easily checked for the nonvacuum
Friedman-Robertson-Walker models of the standard model of
cosmology, and is the cause of the so--called ``horizon problem'' in 
cosmology.  Locality at the singularity can be proved for
Gowdy and polarized $\U(1)$. 
\mnote{where is this done for polarized $\U(1)$?? remove this statement??} 

The picture with bounces and coasting is to a
large extent inspired by the Mixmaster model (Bianchi IX, cf. section
\ref{sec:bianchi}) which is known to have an oscillatory singularity,
cf. Theorem \ref{thm:mixmaster}, and the above scenario 
may therefore be described by the slogan
``generic singularities are spacelike and oscillatory''. 
By an abuse of language
we may say that the oscillatory behavior is of ``mixmaster type''. 

An approach to understanding the asymptotic dynamics near singularities is to
attempt to 
extract, in a systematic way, a dynamical system governing the asymptotic
behavior of the Einstein equations. A promising approach is based on an
analysis of the evolution equations written in 
Hubble--normalized scale--free orthonormal frame variables 
along the
lines discussed for the Gowdy system in section \ref{sec:gowdy}. The general
situation has been discussed in \cite{uggla:etal:G0}. In this setting the
Einstein evolution equations appear in first order form, and the 
natural version of the BKL conjecture is that $\mathbf{\partial_1} U \to 0$
in the direction of the singularity, where $\mathbf{\partial_1}$ is a
scalefree first order operator analogous to the operator 
$e^{-\tau}\partial_x$ for the Gowdy system, and $U$ denotes the
scale free frame variables. 

If this picture is correct, the asymptotic behavior near the singularity can
be understood by analyzing the asymptotic behavior of orbits of the system of
ODE's that one gets by cancelling the terms containing $\mathbf{\partial_1}$
in the first order form of the Einstein evolution equations. The
possibilities that arise have been analyzed in \cite{uggla:etal:G0}. 
It is now important to understand whether or not this version of the BKL
conjecture is correct by doing numerical experiments. No complete results in
this direction have been obtained so far.

The mixmaster type behavior may be
prevented by the presence of certain types of matter such as a scalar field
(stiff fluid). In fact work by Andersson and 
Rendall \cite{andersson:rendall:quiescent}, shows that the
3+1 dimensional Einstein-scalar field equations may be formulated as a
Fuchsian system near the singularity, and AVTD solutions may be constructed
using a singular version of the Cauchy--Kowalevskaya theorem. For the
polarized $\U(1)$ case, the reduced system is 2+1 
dimensional Einstein-scalar field equations. The Fuchsian method has been
applied to this special case by Isenberg and Moncrief
\cite{isenberg:moncrief:U(1)fuchs}. Damour et
al. \cite{damour:etal:kasnerlike} have given a rather
complete study of the application of the Fuchsian
method to spacetimes of general dimension with matter. 

It appears likely that some aspect of the picture sketched above 
will be relevant for the final 
analysis of the large data, global behavior of vacuum spacetimes.

\appendix
\section{Basic Causality concepts}\label{sec:causality}
Here we introduce the basic causality concepts, see 
\cite[Chapter 8]{wald:text},\cite{beem:ehrlich:global,hawking:ellis,penrose:difftop}
for details.  

A compact manifold $\aM$ admits a Lorentz metric if and only if its Euler
characteristic $\chi(\aM)$ vanishes, while all noncompact manifolds admit
Lorentz metrics. 
A vector $V \in T\aM$ is called space--like, null or time--like if $\ame(V,V) >
0, = 0 $ or $< 0$ respectively. $V$ is called causal if it is null or
time--like. A $C^1$ curve $c \subset \aM$ is called time--like (causal) if 
$\dot c$ is time--like (causal). This extends naturally to continuous curves.
A hypersurface $M \subset \aM$ is called acausal if no causal curve meets $M$
more than once, and is called space--like if the  
normal of $M$ is time--like. 

Given an acausal \mnote{achronal or acausal?}
closed subset $S \subset \aM$, the {\bf \index{domain of dependence}{domain
of dependence}}
$D(S)$ is the set of points $p \in \aM$ such that any inextendible
causal curve containing $p$ must intersect $S$. 
If $S \subset \aM$ is a space--like
hypersurface and $D(S) = \aM$, $S$ is called a {\bf \index{Cauchy
surface}{Cauchy surface}}
in $\aM$. If $\aM$ has a Cauchy surface $\aM$ is called {\bf globally
hyperbolic}. A globally hyperbolic spacetime has a global time function, 
i.e. a function $t$ on $\aM$ so that $\anabla t$ is time--like, with the 
level sets  of $t$ being Cauchy surfaces. In particular, $\aM \cong M \times 
\Re$ where $M$ is any Cauchy surface. 

$\aM$ is called {\bf time oriented} if there is a global time--like
vector field on $\aM$. A \index{globally hyperbolic}{globally hyperbolic} 
spacetime is time oriented by
the above, 
and hence it makes sense to talk about the future and past domains of 
dependence $D^+(M)$ and $D^-(M)$ of a Cauchy surface $M$. In a time oriented 
space time, the chronological future $I^+(S)$ of $S \subset \aM$ is the set
of all points reached by future directed time--like curves, of nonzero length, 
starting on $S$. The time--like past $I^-(S)$ is defined
analogously. The causal future and past $J^+(S)$ and $J^-(S)$ is defined
analogously to $I^{\pm}(S)$ with causal curve replacing time--like curve (the
causal curve is allowed to be trivial). 

The variational problem for geodesics (w.r.t. Lorentzian length) is well
behaved precisely when $(\aM, \ame)$ is globally hyperbolic. Let $C(p,q)$ be 
the set of continuous causal curves between $p,q \in \aM$. Then $C(p,q)$ is 
compact w.r.t. uniform convergence for all $p,q \in \aM$, if $(\aM,
\ame)$ is globally hyperbolic. 
This makes global hyperbolicity a natural
assumption in Lorentzian geometry, just as completeness is a natural
assumption in Riemannian geometry. 

Global hyperbolicity may also be characterized by strong causality and
compactness of $C(p,q)$ for all $p,q \in \aM$, or compactness of 
$J^+(p) \cap J^-(q)$ for all $p,q \in \aM$, see \cite[Chapter 8]{wald:text}
for details. 

Given a spacetime $(\aM, \ame)$ and a  space--like hypersurface $M \subset \aM$, 
the future {\bf \index{Cauchy horizon}{Cauchy horizon}} $H^+(M)$ is given by 
$H^+(M) = \overline{D^+(M)} \setminus I^-[D^+(M)]$, with the past Cauchy
horizon $H^-(M)$ defined analogously. The set 
$H(M) = H^+(M) \cup H^-(M)$ is called
the Cauchy horizon of $M$. It can be proved that $H(M) = \partial D(M)$, the
boundary of $D(M)$. 

\bibliographystyle{amsplain}
\bibliography{besse2}

\end{document}

%% file: friedrich.eepic
\setlength{\unitlength}{0.00041667in}
\begingroup\makeatletter\ifx\SetFigFont\undefined
\def\x#1#2#3#4#5#6#7\relax{\def\x{#1#2#3#4#5#6}}%
\expandafter\x\fmtname xxxxxx\relax \def\y{splain}%
\ifx\x\y   
\gdef\SetFigFont#1#2#3{%
  \ifnum #1<17\tiny\else \ifnum #1<20\small\else
  \ifnum #1<24\normalsize\else \ifnum #1<29\large\else
  \ifnum #1<34\Large\else \ifnum #1<41\LARGE\else
     \huge\fi\fi\fi\fi\fi\fi
  \csname #3\endcsname}%
\else
\gdef\SetFigFont#1#2#3{\begingroup
  \count@#1\relax \ifnum 25<\count@\count@25\fi
  \def\x{\endgroup\@setsize\SetFigFont{#2pt}}%
  \expandafter\x
    \csname \romannumeral\the\count@ pt\expandafter\endcsname
    \csname @\romannumeral\the\count@ pt\endcsname
  \csname #3\endcsname}%
\fi
\fi\endgroup
{\renewcommand{\dashlinestretch}{30}
\begin{picture}(5949,5571)(0,-10)
\path(12,162)(2862,4737)(5937,12)
\dottedline{45}(1287,2187)(2937,462)(4512,2187)
\path(1512,3987)(2037,3462)
\blacken\path(1930.934,3525.640)(2037.000,3462.000)(1973.360,3568.066)(1930.934,3525.640)
\path(3612,5337)(2862,4737)
\blacken\path(2936.963,4835.389)(2862.000,4737.000)(2974.445,4788.537)(2936.963,4835.389)
\path(1287,2187)	(1408.230,2171.033)
	(1512.735,2157.390)
	(1602.274,2145.849)
	(1678.605,2136.191)
	(1798.673,2121.647)
	(1887.000,2112.000)

\path(1887,2112)	(1983.374,2102.352)
	(2101.366,2090.840)
	(2233.812,2078.435)
	(2373.544,2066.108)
	(2513.395,2054.829)
	(2646.199,2045.572)
	(2764.790,2039.305)
	(2862.000,2037.000)

\path(2862,2037)	(2959.207,2039.367)
	(3077.791,2045.787)
	(3210.587,2055.233)
	(3350.430,2066.681)
	(3490.157,2079.107)
	(3622.604,2091.485)
	(3740.606,2102.791)
	(3837.000,2112.000)

\path(3837,2112)	(3936.285,2121.917)
	(4071.334,2136.551)
	(4157.213,2146.186)
	(4257.965,2157.660)
	(4375.568,2171.191)
	(4512.000,2187.000)

\put(1212,4062){\makebox(0,0)[lb]{\smash{{{\SetFigFont{6}{7.2}{rm}$\Scri$}}}}}
\put(2937,2187){\makebox(0,0)[lb]{\smash{{{\SetFigFont{6}{7.2}{rm}$M$}}}}}
\put(2637,1437){\makebox(0,0)[lb]{\smash{{{\SetFigFont{6}{7.2}{rm}$D^-(M)$}}}}}
\put(2562,3312){\makebox(0,0)[lb]{\smash{{{\SetFigFont{6}{7.2}{rm}$D^+(M)$}}}}}
\put(3612,5412){\makebox(0,0)[lb]{\smash{{{\SetFigFont{6}{7.2}{rm}$i_+$}}}}}
\put(1212,537){\makebox(0,0)[lb]{\smash{{{\SetFigFont{6}{7.2}{rm}$(\aM, \ame)$}}}}}
\end{picture}
}

%% file: christo.eepic
\setlength{\unitlength}{0.00033333in}
\begingroup\makeatletter\ifx\SetFigFont\undefined
\def\x#1#2#3#4#5#6#7\relax{\def\x{#1#2#3#4#5#6}}%
\expandafter\x\fmtname xxxxxx\relax \def\y{splain}%
\ifx\x\y   
\gdef\SetFigFont#1#2#3{%
  \ifnum #1<17\tiny\else \ifnum #1<20\small\else
  \ifnum #1<24\normalsize\else \ifnum #1<29\large\else
  \ifnum #1<34\Large\else \ifnum #1<41\LARGE\else
     \huge\fi\fi\fi\fi\fi\fi
  \csname #3\endcsname}%
\else
\gdef\SetFigFont#1#2#3{\begingroup
  \count@#1\relax \ifnum 25<\count@\count@25\fi
  \def\x{\endgroup\@setsize\SetFigFont{#2pt}}%
  \expandafter\x
    \csname \romannumeral\the\count@ pt\expandafter\endcsname
    \csname @\romannumeral\the\count@ pt\endcsname
  \csname #3\endcsname}%
\fi
\fi\endgroup
{\renewcommand{\dashlinestretch}{30}
\begin{picture}(12317,8394)(0,-10)
\dashline{60.000}(8858,6012)(12083,8337)
\path(12003.203,8242.489)(12083.000,8337.000)(11968.115,8291.159)
\dashline{60.000}(4133,6987)(4208,87)
\dashline{60.000}(7733,7062)(7733,12)
\path(83,6162)	(256.123,5996.718)
	(381.800,5873.655)
	(533.000,5712.000)

\path(533,5712)	(605.180,5618.653)
	(688.182,5501.052)
	(778.338,5367.249)
	(871.977,5225.295)
	(965.432,5083.242)
	(1055.034,4949.140)
	(1137.113,4831.043)
	(1208.000,4737.000)

\path(1208,4737)	(1313.684,4612.233)
	(1447.089,4463.574)
	(1599.346,4299.364)
	(1761.586,4127.944)
	(1924.941,3957.656)
	(2080.540,3796.842)
	(2219.516,3653.843)
	(2333.000,3537.000)

\path(2333,3537)	(2495.968,3369.528)
	(2597.031,3266.457)
	(2703.976,3158.137)
	(2811.424,3050.301)
	(2913.996,2948.678)
	(3083.000,2787.000)

\path(3083,2787)	(3237.005,2654.224)
	(3361.980,2551.187)
	(3533.000,2412.000)

\path(308,7437)	(441.347,7282.748)
	(556.747,7150.177)
	(656.178,7037.089)
	(741.616,6941.288)
	(878.427,6792.755)
	(983.000,6687.000)

\path(983,6687)	(1069.328,6609.006)
	(1178.189,6517.655)
	(1302.056,6417.374)
	(1433.401,6312.593)
	(1564.696,6207.737)
	(1688.413,6107.234)
	(1797.024,6015.513)
	(1883.000,5937.000)

\path(1883,5937)	(1992.759,5822.670)
	(2121.388,5678.088)
	(2262.524,5513.285)
	(2409.800,5338.290)
	(2556.851,5163.134)
	(2697.312,4997.847)
	(2824.816,4852.459)
	(2933.000,4737.000)

\path(2933,4737)	(3028.702,4644.628)
	(3162.556,4523.212)
	(3248.734,4447.219)
	(3350.382,4358.691)
	(3469.478,4255.870)
	(3608.000,4137.000)

\path(9158,2562)	(9274.477,2705.630)
	(9386.829,2844.201)
	(9495.156,2977.836)
	(9599.556,3106.660)
	(9700.130,3230.796)
	(9796.977,3350.367)
	(9890.196,3465.498)
	(9979.888,3576.311)
	(10066.151,3682.931)
	(10149.086,3785.480)
	(10228.791,3884.083)
	(10305.367,3978.864)
	(10449.529,4157.450)
	(10582.366,4322.227)
	(10704.677,4474.186)
	(10817.257,4614.313)
	(10920.903,4743.598)
	(11016.411,4863.030)
	(11104.579,4973.597)
	(11186.202,5076.289)
	(11333.000,5262.000)

\path(11333,5262)	(11408.503,5357.585)
	(11501.047,5474.551)
	(11604.720,5605.963)
	(11713.606,5744.887)
	(11821.792,5884.388)
	(11923.362,6017.530)
	(12012.403,6137.379)
	(12083.000,6237.000)

\path(12083,6237)	(12163.899,6365.531)
	(12225.412,6469.629)
	(12308.000,6612.000)

\path(8,1287)	(115.918,1292.349)
	(220.030,1297.469)
	(320.427,1302.364)
	(417.203,1307.039)
	(600.262,1315.737)
	(769.947,1323.591)
	(927.001,1330.629)
	(1072.165,1336.877)
	(1206.180,1342.364)
	(1329.789,1347.116)
	(1443.732,1351.162)
	(1548.752,1354.528)
	(1734.986,1359.333)
	(1894.425,1361.751)
	(2033.000,1362.000)

\path(2033,1362)	(2175.622,1359.505)
	(2338.022,1353.877)
	(2517.380,1345.592)
	(2612.535,1340.600)
	(2710.871,1335.123)
	(2812.035,1329.218)
	(2915.675,1322.946)
	(3021.437,1316.365)
	(3128.969,1309.535)
	(3237.918,1302.515)
	(3347.931,1295.364)
	(3458.655,1288.143)
	(3569.739,1280.910)
	(3680.828,1273.724)
	(3791.570,1266.646)
	(3901.613,1259.733)
	(4010.604,1253.047)
	(4118.189,1246.645)
	(4224.017,1240.587)
	(4327.734,1234.933)
	(4428.987,1229.742)
	(4527.424,1225.073)
	(4622.693,1220.986)
	(4802.311,1214.794)
	(4965.021,1211.640)
	(5108.000,1212.000)

\path(5108,1212)	(5275.640,1217.626)
	(5368.259,1222.511)
	(5466.198,1228.646)
	(5569.049,1235.936)
	(5676.401,1244.281)
	(5787.848,1253.584)
	(5902.978,1263.748)
	(6021.384,1274.674)
	(6142.657,1286.266)
	(6266.387,1298.426)
	(6392.165,1311.055)
	(6519.582,1324.057)
	(6648.230,1337.334)
	(6777.699,1350.788)
	(6907.580,1364.321)
	(7037.465,1377.836)
	(7166.943,1391.236)
	(7295.608,1404.422)
	(7423.048,1417.297)
	(7548.855,1429.763)
	(7672.621,1441.723)
	(7793.936,1453.080)
	(7912.392,1463.734)
	(8027.578,1473.590)
	(8139.087,1482.548)
	(8246.508,1490.513)
	(8349.434,1497.385)
	(8447.456,1503.067)
	(8540.163,1507.462)
	(8708.000,1512.000)

\path(8708,1512)	(8825.312,1512.364)
	(8950.333,1511.189)
	(9084.354,1508.393)
	(9228.666,1503.893)
	(9384.561,1497.607)
	(9553.328,1489.452)
	(9736.258,1479.346)
	(9833.439,1473.535)
	(9934.644,1467.206)
	(10040.035,1460.348)
	(10149.775,1452.951)
	(10264.023,1445.004)
	(10382.942,1436.498)
	(10506.693,1427.421)
	(10635.436,1417.763)
	(10769.335,1407.515)
	(10908.549,1396.666)
	(11053.241,1385.206)
	(11203.571,1373.124)
	(11359.701,1360.410)
	(11521.793,1347.054)
	(11690.007,1333.045)
	(11864.506,1318.373)
	(12045.449,1303.028)
	(12138.389,1295.100)
	(12233.000,1287.000)

\path(8933,4062)	(9047.052,4187.711)
	(9157.136,4308.935)
	(9263.351,4425.779)
	(9365.795,4538.347)
	(9464.570,4646.748)
	(9559.774,4751.088)
	(9651.507,4851.472)
	(9739.869,4948.008)
	(9906.877,5129.960)
	(10061.594,5297.794)
	(10204.816,5452.362)
	(10337.341,5594.516)
	(10459.965,5725.107)
	(10573.483,5844.986)
	(10678.694,5955.006)
	(10776.393,6056.016)
	(10952.441,6234.417)
	(11108.000,6387.000)

\path(11108,6387)	(11268.545,6536.069)
	(11370.753,6627.132)
	(11491.981,6733.024)
	(11635.524,6856.600)
	(11804.677,7000.717)
	(11899.888,7081.371)
	(12002.737,7168.232)
	(12113.637,7261.656)
	(12233.000,7362.000)

\put(9533,5187){\makebox(0,0)[lb]{\smash{{{\SetFigFont{8}{9.6}{rm}outgoing gravitational}}}}}
\put(9533,4791){\makebox(0,0)[lb]{\smash{{{\SetFigFont{8}{9.6}{rm}waves}}}}}
\put(5183,1437){\makebox(0,0)[lb]{\smash{{{\SetFigFont{8}{9.6}{rm}$M$}}}}}
\put(4883,7137){\makebox(0,0)[lb]{\smash{{{\SetFigFont{8}{9.6}{rm}inner region}}}}}
\put(1058,7137){\makebox(0,0)[lb]{\smash{{{\SetFigFont{8}{9.6}{rm}outer region}}}}}
\put(11108,8187){\makebox(0,0)[lb]{\smash{{{\SetFigFont{8}{9.6}{rm}$\Scri$}}}}}
\end{picture}
}

%% file: ckn.eepic
\setlength{\unitlength}{0.00033333in}
\begingroup\makeatletter\ifx\SetFigFont\undefined
\def\x#1#2#3#4#5#6#7\relax{\def\x{#1#2#3#4#5#6}}%
\expandafter\x\fmtname xxxxxx\relax \def\y{splain}%
\ifx\x\y   
\gdef\SetFigFont#1#2#3{%
  \ifnum #1<17\tiny\else \ifnum #1<20\small\else
  \ifnum #1<24\normalsize\else \ifnum #1<29\large\else
  \ifnum #1<34\Large\else \ifnum #1<41\LARGE\else
     \huge\fi\fi\fi\fi\fi\fi
  \csname #3\endcsname}%
\else
\gdef\SetFigFont#1#2#3{\begingroup
  \count@#1\relax \ifnum 25<\count@\count@25\fi
  \def\x{\endgroup\@setsize\SetFigFont{#2pt}}%
  \expandafter\x
    \csname \romannumeral\the\count@ pt\expandafter\endcsname
    \csname @\romannumeral\the\count@ pt\endcsname
  \csname #3\endcsname}%
\fi
\fi\endgroup
{\renewcommand{\dashlinestretch}{30}
\begin{picture}(13313,10038)(0,-10)
\path(12,4308)(5112,4308)
\path(6537,4308)(12012,4308)
\drawline(6462,4533)(6462,4533)
\drawline(6462,4533)(6462,4533)
\dottedline{45}(8862,4458)(11862,8133)
\dottedline{45}(10212,4833)(12537,7833)
\dottedline{45}(9237,3933)(12312,558)
\dottedline{45}(10662,3633)(12537,1908)
\dottedline{45}(3312,4458)(237,7683)
\dottedline{45}(1812,4533)(87,6558)
\dottedline{45}(3012,3933)(462,1308)
\dottedline{45}(1962,3783)(237,1983)
\dottedline{45}(9612,4683)(12162,7983)
\blacken\path(12112.365,7869.702)(12162.000,7983.000)(12064.888,7906.389)(12112.365,7869.702)
\path(987,3708)(1137,4308)
\blacken\path(1137.000,4184.307)(1137.000,4308.000)(1078.791,4198.859)(1137.000,4184.307)
\path(4662,3633)(4887,4308)
\blacken\path(4877.513,4184.671)(4887.000,4308.000)(4820.592,4203.645)(4877.513,4184.671)
\path(11112,3858)(10887,4308)
\blacken\path(10967.498,4214.085)(10887.000,4308.000)(10913.833,4187.252)(10967.498,4214.085)
\path(4662,6033)(5712,5958)
\blacken\path(5590.168,5936.626)(5712.000,5958.000)(5594.442,5996.473)(5590.168,5936.626)
\path(4662,6108)(5112,6783)
\blacken\path(5070.397,6666.513)(5112.000,6783.000)(5020.474,6699.795)(5070.397,6666.513)
\path(7282,7538)(6082,6488)
\blacken\path(6152.554,6589.598)(6082.000,6488.000)(6192.064,6544.443)(6152.554,6589.598)
\dashline{180.000}(312,8433)(4212,4308)
\dashline{180.000}(7962,4383)(11862,9333)
\dashline{180.000}(4137,4233)(1137,183)
\dashline{180.000}(7962,4233)(11112,33)
\path(3312,1383)(2712,2133)
\blacken\path(2810.389,2058.037)(2712.000,2133.000)(2763.537,2020.555)(2810.389,2058.037)
\path(4887,1383)(9162,2358)
\blacken\path(9051.675,2302.068)(9162.000,2358.000)(9038.333,2360.566)(9051.675,2302.068)
\path(7137,5058)(6387,4458)
\path(6461.963,4556.389)(6387.000,4458.000)(6499.445,4509.537)
\path(5187,4758)(5637,5733)
\path(5637,4758)(5862,5583)
\path(4662,10008)	(4668.084,9864.294)
	(4674.242,9725.652)
	(4680.489,9591.950)
	(4686.839,9463.064)
	(4693.305,9338.872)
	(4699.901,9219.249)
	(4706.641,9104.071)
	(4713.538,8993.216)
	(4720.606,8886.560)
	(4727.860,8783.978)
	(4735.311,8685.348)
	(4742.975,8590.546)
	(4758.996,8411.930)
	(4776.030,8247.143)
	(4794.188,8095.194)
	(4813.581,7955.096)
	(4834.317,7825.860)
	(4856.507,7706.497)
	(4880.260,7596.019)
	(4905.687,7493.435)
	(4962.000,7308.000)

\path(4962,7308)	(5009.802,7184.808)
	(5073.055,7049.791)
	(5154.615,6898.337)
	(5257.339,6725.831)
	(5317.530,6630.242)
	(5384.083,6527.659)
	(5457.355,6417.506)
	(5537.703,6299.205)
	(5625.485,6172.182)
	(5721.057,6035.857)
	(5824.776,5889.656)
	(5937.000,5733.000)

\path(6987,10008)	(6985.923,9884.382)
	(6984.681,9765.116)
	(6983.268,9650.097)
	(6981.678,9539.216)
	(6979.902,9432.368)
	(6977.935,9329.447)
	(6975.769,9230.345)
	(6973.398,9134.957)
	(6968.011,8954.897)
	(6961.721,8788.413)
	(6954.471,8634.655)
	(6946.208,8492.771)
	(6936.875,8361.910)
	(6926.418,8241.221)
	(6914.783,8129.851)
	(6901.913,8026.950)
	(6872.254,7843.147)
	(6837.000,7683.000)

\path(6837,7683)	(6795.206,7537.731)
	(6738.484,7375.750)
	(6664.196,7191.345)
	(6619.642,7088.948)
	(6569.708,6978.802)
	(6514.063,6860.194)
	(6452.380,6732.410)
	(6384.328,6594.734)
	(6309.578,6446.453)
	(6227.799,6286.854)
	(6138.663,6115.221)
	(6041.840,5930.841)
	(5990.443,5833.648)
	(5937.000,5733.000)

\path(5862,9858)	(5850.513,9724.842)
	(5840.062,9600.992)
	(5830.617,9485.983)
	(5822.152,9379.350)
	(5814.640,9280.623)
	(5808.052,9189.338)
	(5797.541,9027.221)
	(5790.400,8889.265)
	(5786.410,8771.732)
	(5785.350,8670.889)
	(5787.000,8583.000)

\path(5787,8583)	(5800.849,8410.567)
	(5813.252,8309.880)
	(5828.497,8201.564)
	(5845.977,8087.134)
	(5865.085,7968.104)
	(5885.212,7845.989)
	(5905.751,7722.304)
	(5926.095,7598.562)
	(5945.636,7476.278)
	(5963.767,7356.967)
	(5979.880,7242.142)
	(5993.367,7133.319)
	(6003.621,7032.011)
	(6012.000,6858.000)

\path(6012,6858)	(5995.174,6673.691)
	(5972.134,6529.046)
	(5956.189,6438.217)
	(5937.000,6333.000)

\path(5112,4308)	(5262.000,4308.000)

\path(5262,4308)	(5412.000,4383.000)

\path(5412,4383)	(5562.000,4233.000)

\path(5562,4233)	(5683.028,4351.646)
	(5787.000,4458.000)

\path(5787,4458)	(5865.705,4304.786)
	(5937.000,4158.000)

\path(5937,4158)	(6045.345,4305.585)
	(6162.000,4458.000)

\path(6162,4458)	(6234.015,4348.091)
	(6312.000,4233.000)

\path(6312,4233)	(6387.000,4308.000)

\path(6387,4308)	(6537.000,4308.000)

\path(6348,4846)(6048,5671)
\put(2862,7608){\makebox(0,0)[lb]{\smash{{{\SetFigFont{7}{8.4}{rm}$(\aM, \ame)$}}}}}
\put(612,5358){\makebox(0,0)[lb]{\smash{{{\SetFigFont{7}{8.4}{rm}$D(M\backslash K)$}}}}}
\put(312,3408){\makebox(0,0)[lb]{\smash{{{\SetFigFont{7}{8.4}{rm}$M\backslash K$}}}}}
\put(4437,3333){\makebox(0,0)[lb]{\smash{{{\SetFigFont{7}{8.4}{rm}$K$}}}}}
\put(3612,5733){\makebox(0,0)[lb]{\smash{{{\SetFigFont{7}{8.4}{rm}horizon}}}}}
\put(7362,7533){\makebox(0,0)[lb]{\smash{{{\SetFigFont{7}{8.4}{rm}singularity}}}}}
\put(7212,5058){\makebox(0,0)[lb]{\smash{{{\SetFigFont{7}{8.4}{rm}strong field}}}}}
\put(11187,3483){\makebox(0,0)[lb]{\smash{{{\SetFigFont{7}{8.4}{rm}$M\backslash K$}}}}}
\put(12237,8058){\makebox(0,0)[lb]{\smash{{{\SetFigFont{7}{8.4}{rm}to $\Scri$}}}}}
\put(3387,1158){\makebox(0,0)[lb]{\smash{{{\SetFigFont{7}{8.4}{rm}$H(M\backslash K)$}}}}}
\end{picture}
}

%% file: am.eepic
\setlength{\unitlength}{0.00041667in}
\begingroup\makeatletter\ifx\SetFigFont\undefined
\def\x#1#2#3#4#5#6#7\relax{\def\x{#1#2#3#4#5#6}}%
\expandafter\x\fmtname xxxxxx\relax \def\y{splain}%
\ifx\x\y   
\gdef\SetFigFont#1#2#3{%
  \ifnum #1<17\tiny\else \ifnum #1<20\small\else
  \ifnum #1<24\normalsize\else \ifnum #1<29\large\else
  \ifnum #1<34\Large\else \ifnum #1<41\LARGE\else
     \huge\fi\fi\fi\fi\fi\fi
  \csname #3\endcsname}%
\else
\gdef\SetFigFont#1#2#3{\begingroup
  \count@#1\relax \ifnum 25<\count@\count@25\fi
  \def\x{\endgroup\@setsize\SetFigFont{#2pt}}%
  \expandafter\x
    \csname \romannumeral\the\count@ pt\expandafter\endcsname
    \csname @\romannumeral\the\count@ pt\endcsname
  \csname #3\endcsname}%
\fi
\fi\endgroup
{\renewcommand{\dashlinestretch}{30}
\begin{picture}(5124,5889)(0,-10)
\put(2337,1362){\ellipse{150}{150}}
\put(2862,1212){\ellipse{150}{150}}
\dottedline{45}(1962,2787)(2262,1737)
\dottedline{45}(2712,2862)(2637,1512)
\dottedline{45}(3312,2562)(3012,1512)
\path(2637,12)(12,5562)
\path(2637,12)(5112,5862)
\dottedline{45}(1662,3987)(1062,5787)
\dottedline{45}(2862,3687)(3087,5712)
\drawline(3837,3912)(3837,3912)
\dottedline{45}(3762,3987)(4287,5787)
\path(1287.000,2937.000)	(1256.531,3019.031)
	(1240.125,3115.125)
	(1237.781,3225.281)
	(1249.500,3349.500)

\path(1249.500,3349.500)	(1282.312,3478.406)
	(1343.250,3602.625)
	(1432.312,3722.156)
	(1549.500,3837.000)

\path(1549.500,3837.000)	(1676.062,3937.781)
	(1793.250,4015.125)
	(1901.062,4069.031)
	(1999.500,4099.500)

\path(1999.500,4099.500)	(2097.938,4104.188)
	(2205.750,4080.750)
	(2322.938,4029.188)
	(2449.500,3949.500)

\path(2449.500,3949.500)	(2573.719,3862.781)
	(2683.875,3790.125)
	(2779.969,3731.531)
	(2862.000,3687.000)

\path(2862.000,3687.000)	(2941.688,3663.562)
	(3030.750,3668.250)
	(3129.188,3701.062)
	(3237.000,3762.000)

\path(3237.000,3762.000)	(3347.156,3829.969)
	(3452.625,3883.875)
	(3553.406,3923.719)
	(3649.500,3949.500)

\path(3649.500,3949.500)	(3736.219,3954.188)
	(3808.875,3930.750)
	(3912.000,3799.500)

\path(3912.000,3799.500)	(3944.812,3705.750)
	(3968.250,3612.000)
	(3982.312,3518.250)
	(3987.000,3424.500)

\path(3987.000,3424.500)	(3977.625,3326.062)
	(3949.500,3218.250)
	(3902.625,3101.062)
	(3837.000,2974.500)

\path(3837.000,2974.500)	(3759.656,2852.625)
	(3677.625,2749.500)
	(3590.906,2665.125)
	(3499.500,2599.500)

\path(3499.500,2599.500)	(3401.062,2559.656)
	(3293.250,2552.625)
	(3176.062,2578.406)
	(3049.500,2637.000)

\path(3049.500,2637.000)	(2927.625,2709.656)
	(2824.500,2777.625)
	(2740.125,2840.906)
	(2674.500,2899.500)

\path(2674.500,2899.500)	(2543.250,2974.500)
	(2463.562,2983.875)
	(2374.500,2974.500)

\path(2374.500,2974.500)	(2283.094,2953.406)
	(2196.375,2927.625)
	(2114.344,2897.156)
	(2037.000,2862.000)

\path(2037.000,2862.000)	(1962.000,2824.500)
	(1887.000,2787.000)
	(1812.000,2749.500)
	(1737.000,2712.000)

\path(1737.000,2712.000)	(1596.375,2665.125)
	(1474.500,2674.500)

\path(1474.500,2674.500)	(1390.125,2712.000)
	(1362.000,2749.500)

\path(1362.000,2749.500)	(1362.000,2777.625)
	(1362.000,2787.000)

\path(1362.000,2787.000)	(1343.250,2824.500)
	(1287.000,2937.000)

\path(1662,3237)	(1742.872,3385.609)
	(1770.298,3473.187)
	(1812.000,3537.000)

\path(1812,3537)	(1953.720,3603.945)
	(2036.941,3618.270)
	(2112.000,3612.000)

\path(2112,3612)	(2201.730,3550.466)
	(2262.000,3462.000)

\path(2262,3462)	(2288.584,3351.375)
	(2262.000,3237.000)

\path(2262,3237)	(2115.750,3168.030)
	(2031.521,3161.918)
	(1962.000,3162.000)

\path(1962,3162)	(1882.908,3170.819)
	(1782.727,3191.767)
	(1662.000,3237.000)

\path(1662,3237)	(1662.000,3237.000)

\path(3162,3237)	(3242.430,3347.659)
	(3312.000,3462.000)

\path(3312,3462)	(3413.573,3529.399)
	(3537.000,3537.000)

\path(3537,3537)	(3621.750,3399.585)
	(3624.947,3313.673)
	(3612.000,3237.000)

\path(3612,3237)	(3550.466,3147.270)
	(3462.000,3087.000)

\path(3462,3087)	(3387.000,3087.000)

\path(3387,3087)	(3309.142,3119.363)
	(3237.000,3162.000)

\path(3237,3162)	(3162.000,3237.000)

\path(3162,3237)	(3162.000,3237.000)

\path(2077,1151)	(2051.624,1260.905)
	(2002.000,1376.000)

\path(2002,1376)	(2034.624,1492.000)
	(2095.360,1612.092)
	(2184.416,1707.889)
	(2302.000,1751.000)

\path(2302,1751)	(2390.972,1688.540)
	(2452.000,1601.000)

\path(2452,1601)	(2524.142,1558.366)
	(2602.000,1526.000)

\path(2602,1526)	(2685.920,1517.187)
	(2790.655,1518.166)
	(2894.812,1523.062)
	(2977.000,1526.000)

\path(2977,1526)	(3049.345,1541.469)
	(3127.000,1526.000)

\path(3127,1526)	(3153.992,1409.994)
	(3127.000,1301.000)

\path(3127,1301)	(3111.063,1212.893)
	(3085.622,1105.516)
	(3043.370,1002.131)
	(2977.000,926.000)

\path(2977,926)	(2863.000,902.041)
	(2752.000,926.000)

\path(2752,926)	(2669.995,996.395)
	(2602.000,1076.000)

\path(2602,1076)	(2527.000,1151.000)

\path(2527,1151)	(2408.294,1123.179)
	(2302.000,1076.000)

\path(2302,1076)	(2227.971,1068.080)
	(2152.000,1076.000)

\path(2152,1076)	(2077.000,1151.000)

\path(2077,1151)	(2077.000,1151.000)

\put(2487,3387){\makebox(0,0)[lb]{\smash{{{\SetFigFont{9}{10.8}{rm}$M$}}}}}
\put(1887,5187){\makebox(0,0)[lb]{\smash{{{\SetFigFont{9}{10.8}{rm}$(\aM, \ame)$}}}}}
\end{picture}
}